\documentclass[aps,prd,superscriptaddress,showpacs,showkeys,a4paper,superscriptaddress,10pt,nofootinbib,notitlepage]{revtex4-1}

\usepackage[utf8]{inputenc}
\usepackage[english]{babel}
\usepackage{amsmath}
\usepackage{amssymb}
\usepackage{amsfonts}
\usepackage{graphicx}
\usepackage{bm}
\usepackage{cancel}
\usepackage{color}
\usepackage{txfonts}
\usepackage[T1]{fontenc}

\usepackage[colorlinks,citecolor=blue,linkcolor=blue,urlcolor=blue]{hyperref}
\usepackage[tiny,center,uppercase]{titlesec}
\usepackage[a4paper,textwidth=380pt,textheight=592pt,top=107pt,left=80pt]{geometry}

\renewcommand{\vec}{\boldsymbol}
\newcommand{\la}{\langle}
\newcommand{\ra}{\rangle}

\newcommand{\thp}[2]{\vec #1\cdot\vec #2}
\newcommand{\opA}[1]{\boldsymbol{\mathsf{#1}}}
\newcommand{\opa}[1]{\mathsf{#1}}
\newcommand\ri{\mathrm{i}}

\allowdisplaybreaks[2]

\begin{document}
	
\title{Regularization of ultraviolet divergence for a particle interacting with a scalar quantum field}

\author{O.\ D.\ Skoromnik}
\email[Corresponding author: ]{olegskor@gmail.com}
\affiliation{Max Planck Institute for Nuclear Physics, Saupfercheckweg 1, 69117 Heidelberg, Germany}
\author{I.\ D.\ Feranchuk}
\affiliation{Belarusian State University, 4 Nezavisimosty Ave., 220030, Minsk,   Belarus}
\author{D. V. Lu}
\affiliation{Belarusian State University, 4 Nezavisimosty Ave., 220030, Minsk,   Belarus}
\author{C. H. Keitel}
\affiliation{Max Planck Institute for Nuclear Physics, Saupfercheckweg 1, 69117 Heidelberg, Germany}

\begin{abstract}
When a non-relativistic particle interacts with a scalar quantum field, the standard perturbation theory leads to a dependence of the energy of its ground state on an undefined parameter---``momentum cut-off''---due to the ultraviolet divergence. We show that the use of non-asymptotic states of the system results in a calculation scheme in which all observable quantities remain finite and continuously depend on the coupling constant without any additional parameters. It is furthermore demonstrated that the divergence of traditional perturbation series is caused by the energy being a function with a logarithmic singularity for small values of the coupling constant.
\end{abstract}

\maketitle

\section{Introduction} 
\label{sec:introduction}
A characteristic property of the majority of quantum field theories (QFT) is the divergence of integrals appearing in the perturbation theory for the calculation of physical quantities such as mass and charge of the interacting particles. The divergences appear in the integrations over momenta in intermediate states both on the lower limit (infrared divergence) and on the upper limit (ultraviolet divergence). In order to circumvent this difficulty, a renormalization procedure is used, which allows the redefinition of the initial parameters of the system through their observable values. The renormalization scheme was firstly developed for quantum electrodynamics (QED) \cite{PhysRev.75.1736,*PhysRev.85.631,*PhysRev.95.1300,*Stueckelberg1953} and later generalized to other QFT models \cite{'tHooft1971173,*Hooft1971167,*'tHooft1972189}. These schemes can be used for so-called renormalizable theories, for which the reconstructed perturbation theory can be built in a way that the infinite values are included in the definition of ``physical'' charge and mass and, therefore, do not appear in other observables of the system. However, even the founders of QED anticipated ``that the renormalization theory is simply a way to sweep the difficulties of the divergences in electrodynamics under the rug.'' (R. P. Feynman \cite{Feynman12081966}). In many papers P. A. M. Dirac wrote that this approach was in contradiction with logical principles of quantum mechanics \cite{Dirac1981,Dirac1981book}.

Accordingly, the question arises whether these divergences are an intrinsic property of quantum field models or they are caused by the application of perturbation theory for the calculation of physical quantities, which are non-analytical functions of the coupling constant such as in the theory of superconductivity \cite{PhysRev.108.1175}. A large number of works is devoted to this problem, nevertheless a solution still has not been found up to now. However, this question is of great importance for a correct mathematical formulation of fundamental physical theories and is essential for examining the applicability of non-renormalizable theories \cite{PhysRevD.88.125014,*PhysRevD.87.065024} for the description of real physical systems.

Let us recall that in standard perturbation theory the Hamiltonian of non-interacting fields is used as a zeroth-order approximation, while Fock states of the free fields are employed for the calculation of the transition matrix elements in the subsequent corrections to observable characteristics of the system. This approach is based on the assumption of an asymptotic switch off of the interaction between the fields \cite{LandauQED}. However, in a series of works it has been shown \cite{PhysRev.140.B1110,Faddeev1970,PhysRevD.11.3481,PhysRev.173.1527,*PhysRev.174.1882,*PhysRev.175.1624} that the infrared divergence arises just because of the use of asymptotically free field states. As follows from reference \cite{PhysRev.140.B1110} and the subsequent publication \cite{Faddeev1970}, the infrared divergence disappears in all orders of perturbation theory in QED if, in the zeroth-order approximation, the coherent states of the electromagnetic field bound to the particle are used and the parameters of these states are appropriately chosen.

At first glance, this may contradict representation theory in quantum mechanics, in accordance to which the result of the calculation of the observable characteristics of the system should not depend on the choice of basis states, provided that those form a full basis in a Hilbert space as for free field states. However, this statement is correct only for the exact solution of the problem, whereas individual terms of the perturbation series can change with a different choice of basis in zeroth-order approximation. As was demonstrated in reference \cite{Feranchuk2015,Feranchuk1995370} the transition from one basis to another corresponds to the partial summation of a divergent series within standard perturbation theory  and allows for the non-perturbative calculation of subsequent corrections in the form of a convergent sequence. A good example of how the basis choice influences the approximate calculation of the characteristics of the quantum system with a continuous spectrum is given by the scattering at a Coulomb potential \cite{PhysRevA.82.052703,*PhysRevA.70.052701}. In this well known case, the wave function of the system has no singularities. In contrast, Born's scattering amplitude, approximately calculated via an asymptotically free basis, displays a singularity for scattering at small angles. As was demonstrated in references \cite{PhysRevA.82.052703,*PhysRevA.70.052701}, this singularity does not appear with the use of non-asymptotic wave functions.

The main goal of our work is to investigate whether a proper choice of basis in zeroth-order approximation allows to construct a calculation scheme free of ultraviolet divergences. In order not to overload the proposed approach with details related to the internal degrees of freedom and to render all calculations as transparent as possible, we investigate as a representative example a model system, which consists of a non-relativistic particle without spin interacting with a scalar quantum field. A standard-perturbation-theory series, in this case, does not exhibit infrared divergence, contains however ultraviolet divergence. This results in a dependence of the energy of the ground state on the undefined momentum cut-off, which is required for the calculation of high-order corrections. Consequently, our task is to prove that the energy of the ground state of the considered model system can be calculated without any additional parameters such as a momentum cut-off. At the same time, it is important to show that the energy of the system is a non-analytical function of the coupling constant and consequently can not be represented as a series in the framework of conventional perturbation theory.

With the inclusion of a field polarization our employed model coincides with non-relativistic QED \cite{Healy1982} or if the field is scalar it has a physical realization in solids \cite{Toyozawa01071961}, where however, due to the discrete structure of a crystal, a natural momentum cut-off intrinsically appears, defined via the Brillouin-zone boundary. In free space this regularization is not present and has to be artificially included, e.g., via lattice models \cite{Smit2002}, where the boundary momentum is defined through an artificial lattice period. In contrast, in our formulation we will consider a system in free space without neither natural nor artificial cut-off.

In addition, in a series of works \cite{Spohn1998,*Spohn1989,Amann1991414,*Arai1997455,jmp.5.1190.1964,Bach1998299,*Bach2007426,*Chen20082555} an analogous model of a particle interacting with a scalar quantum field with the momentum cut-off was used for the investigation of the fundamental mathematical problem of the existence of the solutions of the Schr\"{o}dinger equation.

The article is organized in the following way. In section \ref{sec:model_description} the model of a non-relativistic particle without spin interacting with a scalar quantum field is described and its parameters
are calculated in the framework of conventional perturbation theory.
In section \ref{sec:iteration_scheme_basis_zero_order_approximation_to_the_energy_of_the_system}
the basis of non-asymptotic states is investigated and the iteration scheme of the calculations is presented.
The zeroth-order approximation, which is found to be free of ultraviolet divergence for the energy and effective mass is then worked out using this basis.
In section \ref{sec:second_order_iteration_for_the_energy_convergence}
the proposed iteration scheme is employed for computing the correction to the zeroth-order approximation of the energy. The convergence of all integrals is demonstrated and the character of the singularity
of the energy as a non-analytic function of the coupling constant is determined in the weak coupling limit. In addition, the details of all calculations are presented in appendices.

\section{Model description} 
\label{sec:model_description}
Let us examine the Hamiltonian of the system consisting of a non-relativistic particle interacting with a scalar quantum field
 \begin{align}
	\opa H &= \opa H_0+\opa H_{\text{int}}, \label{eq:model_description1}
	\\
	\opa H_0 &= \frac{\opA p^2}{2}+\sum_{\vec k}\omega_{\vec k} \opa a^\dag_{\vec k}\opa a_{\vec k}, \label{eq:model_description2}
	\\
	\opa H_{\text{int}} &=\frac{g}{\sqrt{2\Omega}}\sum_{\vec k}A_{\vec k} \left(e^{\ri\thp{k}{r}}\opa a_{\vec k}+e^{-\ri\thp{k}{r}}\opa a^\dag_{\vec k}\right). \label{eq:model_description3}
\end{align}

Here, we select the system of units in which $m = 1$, $\hbar = c = 1$, the momentum operator $\opA p=-\ri \nabla$, normalization volume $\Omega$, vertex function $A_{\vec k} = 1/{\sqrt{\omega_{\vec k}}}$, creation (annihilation) operators $\opa a_{\vec k}^\dag$ ($\opa a_{\vec k}$) of the field mode with the frequency $\omega_{\vec k}=k=\vert{\vec k}\vert$, and the coupling constant $g$. The real physical system, which is described via Hamiltonian (\ref{eq:model_description1}) corresponds to an electron interacting with acoustic phonons in a continuous model of a crystal \cite{Toyozawa01071961}. If we choose $\omega_{\vec k} = 1$, $A_{\vec k} = 1/k$, and $g = \sqrt{8\pi \alpha}$, operator (\ref{eq:model_description1}) corresponds to the Fr\"ohlich Hamiltonian \cite{Froehlich1954}, which describes the interaction of an electron with optical phonons in a crystal, i.e., the so-called ``polaron'' problem \cite{RevModPhys.63.63,*Mitra198791,PhysRev.97.660,Spohn1987278}.

The total momentum operator
\begin{align}
	\opA P = -\ri \nabla + \sum_{\vec k}\vec k\opa a^\dag_{\vec k}\opa a_{\vec k}\label{eq:model_description4}
\end{align}
commutes with the Hamiltonian of the system (\ref{eq:model_description1}) and consequently the eigenvalues $E(\vec P)$ and eigenfunctions $|\Psi_{\vec P}\ra$ are defined as solutions of the following system of equations
\begin{align}
	\opa H|\Psi_{\vec P}\ra &= E(\vec P)|\Psi_{\vec P}\ra, \label{eq:model_description5}
	\\
	\opA P|\Psi_{\vec P}\ra &= \vec P |\Psi_{\vec P}\ra. \label{eq:model_description6}
\end{align}

In the conventional perturbation expansion over the coupling constant in the zeroth-order approximation the solution of the stationary Schr\"odinger equation with Hamiltonian (\ref{eq:model_description2}) is simply determined and corresponds to the free particle with momentum $\vec p$ and Fock states of the phonon field with the set of occupation numbers $\{n_{\vec k_1},n_{\vec k_2},\ldots\}\equiv \{n_{\vec k}\}$:
\begin{align}
	|\Psi^{(0)}_{\vec p,\{n_{\vec k}\}}\ra &= \frac{e^{\ri \thp{p}{r}}}{\sqrt{\Omega}}|\{n_{\vec k}\}\ra,\quad \sum_{\vec k}  \opa a^\dag_{\vec k}\opa a_{\vec k}|\{ n_{\vec k}\}\ra =\sum_{\vec k}n_{\vec k} |\{ n_{\vec k}\}\ra, \label{eq:model_description7}
	\\
	E^{(0)}(\vec P, \{ n_{\vec k}\}) &= \frac{1}{2}\left(\vec P - \sum_{\vec k}\vec k n_{\vec k}\right)^2 + \sum_{\vec k}\omega_{\vec k} n_{\vec k},\quad \vec P = \vec p + \sum_{\vec k}\vec k n_{\vec k}. \label{eq:model_description8}
\end{align}

Let us suppose that the system is in the ground state of the phonon field $\{n_{\vec k}\} = 0$, which leads to the following eigenfunction and eigenvalue
\begin{align}
	|\Psi^{(0)}_{\vec P, 0}\ra &=  \frac{e^{\ri \thp{P}{r}}}{\sqrt{\Omega}} |0\ra \label{eq:model_description9}
	\\
	E^{(0)}(\vec P,0) &= \frac{P^2}{2}, \quad \vec P = \vec p. \label{eq:model_description10}
\end{align}

The first non-vanishing correction to the system energy arises in the second order of perturbation theory (single-phonon intermediate transitions) and corresponds to the self-energy diagram, which defines the mass operator $\Sigma(\vec P)$ and is determined as
\begin{align}
	\label{eq:model_description11}
	\Sigma (\vec P) &= \Delta E^{(2)}(\vec P,0) = - \frac{g^2}{2 \Omega} \sum_{\vec k} \frac{1}{\omega_{\vec k}} \frac{1}{k^2/2 - \thp{P}{k} + \omega_k} =  -   \frac{g^2}{16 \pi^3} \int \frac{d \vec k}{k [k^2/2 - \thp{P}{k} + k]}.
\end{align}

In order to select bound state energy $E_b = E^{(2)}(0,0)$ and effective mass $m^*$ of a particle we expand the energy in a series over $\vec P$ up to second order
\begin{align}
	E^{(0)}(\vec P,0) + \Delta E^{(2)}(\vec P,0) \approx   E_b + \frac{P^2}{2 m^*} \equiv    -   \frac{g^2}{16 \pi^3} \int \frac{d \vec k}{k [k^2/2  + k]} + \frac{P^2}{2} - \frac{g^2}{16 \pi^3} \int \frac{d \vec k}{k [k^2/2  + k]^3}(\thp{P}{k})^2.\label{eq:model_description12}
\end{align}

The first integral in equation (\ref{eq:model_description12}) logarithmically diverges, that is the bound state energy depends on the momentum cut-off $K$
\begin{align}
	E_b = - \frac{g^2}{2 \pi^2}\ln \left(\frac{K}{2} +1\right),\label{eq:model_description13}
\end{align}
and becomes infinite when $K\rightarrow\infty$, such that the correction to the energy is undefined in the framework of the perturbation theory for our model \cite{Messiah1981}.

At the same time, the corrected mass is well defined and equal to
\begin{align}
	\frac{1}{m^*} \simeq 1  - \frac{g^2}{6 \pi^2}; \quad m^* \simeq 1  + \frac{g^2}{6 \pi^2 }. \label{eq:model_description14}
\end{align}

In contrast to this  in the polaron problem all integrals are convergent because they contain  in the denominator the additional power of $k$. The corresponding quantities for the polaron problem read as
\cite{RevModPhys.63.63,*Mitra198791,PhysRev.97.660,Spohn1987278}
\begin{align}
	E_b \simeq - \alpha; \quad m^* \simeq 1  + \frac{\alpha}{6 }.\label{eq:model_description15}
\end{align}

It is important to stress here that in our model, the interaction energy between particle and field is observable and consequently, the infinite energy (\ref{eq:model_description12}) can not be included in the mass renormalization. Thus, we can conclude that the use of perturbation theory for two physically close quantum-field models leads to qualitatively different results. Therefore, a modification of the calculation method of subsequent corrections to the energy for our model is required and appears achievable.

\section{Iteration scheme, basis choice and zeroth-order approximation of the system's energy} 
\label{sec:iteration_scheme_basis_zero_order_approximation_to_the_energy_of_the_system}
In order to build an iteration scheme not in the framework of perturbation theory we will employ the operator method (OM) for the solution of the Schr\"odinger equation, which was introduced in reference \cite{Feranchuk1982211} and its detailed explanation is given in the monograph \cite{Feranchuk2015,Feranchuk1995370}. Let us quickly revise here the basics of this method. Suppose, the eigenvalues $E_\mu$ and eigenvectors $|\Psi_{\mu}\ra$ with a set of quantum numbers $\mu$ of the stationary Schr\"odinger equation need to be found:
\begin{align}
	\opa H |\Psi_{\mu}\ra = E_{\mu}|\Psi_{\mu}\ra.\label{eq:iteration_scheme_basis_zero_order_approximation_to_the_energy_of_the_system_1}
\end{align}

In contrast to perturbation theory, where the Hamiltonian $\opa H$ of the system is split into the zeroth-order approximation and perturbation parts, according to OM the total Hamiltonian is taken into account as is, while however, the state vector is probed via an approximate state:
\begin{align*}
	|\Psi_{\mu}\ra \approx |\psi_{\mu}(\omega_{\mu})\ra,
\end{align*}
which depends on a set of variational parameters $\omega_{\mu}$. Then, the exact solution can be represented as a series
\begin{align}
	|\Psi_{\mu}\ra = |\psi_{\mu}(\omega_{\mu})\ra + \sum_{\nu \neq \mu} C_{\mu \nu}|\psi_{\nu} (\omega_{\mu} )\ra. \label{eq:iteration_scheme_basis_zero_order_approximation_to_the_energy_of_the_system_2}
\end{align}
Here we want to pay attention to the fact that for a given set of quantum numbers $\mu$, the set $\omega_{\mu}$ is fixed. By plugging the expansion (\ref{eq:iteration_scheme_basis_zero_order_approximation_to_the_energy_of_the_system_2}) into Schr\"odinger's equation (\ref{eq:iteration_scheme_basis_zero_order_approximation_to_the_energy_of_the_system_1}) and projecting on different states $|\psi_{\mu}(\omega_{\mu})\ra$ and $|\psi_{\nu}(\omega_{\mu})\ra$ one obtains the equations for the energies $E_{\mu}$ and coefficients $C_{\mu \nu}$:
\begin{align}
	E_{\mu} &= \left[1 +   \sum_{\nu \neq \mu} C_{\mu \nu}I_{\mu \nu}\right]^{-1} \left[H_{\mu \mu} + \sum_{\nu \neq \mu} C_{\mu \nu}H_{\mu \nu}\right]; \label{eq:iteration_scheme_basis_zero_order_approximation_to_the_energy_of_the_system_3}
	\\
	C_{\mu \gamma} &= \left[E_{\mu} - H_{\gamma \gamma}\right]^{-1}\left[ H_{\gamma \mu} - E_{\mu}I_{\gamma \mu} + \sum_{\nu \neq \mu \neq \gamma}C_{\mu \nu} (H_{\gamma \nu} - E_{\mu}I_{\gamma \nu})\right]; \label{eq:iteration_scheme_basis_zero_order_approximation_to_the_energy_of_the_system_4}
	\\
	H_{\mu \nu} &\equiv \la\psi_{\mu} (\omega_{\mu} )| \opa H | \psi_{\nu} (\omega_{\mu})\ra; \quad I_{\mu \nu} \equiv \la\psi_{\mu} (\omega_{\mu} )|  \psi_{\nu} (\omega_{\mu})\ra. \nonumber
\end{align}

It is important to stress here that all matrix elements are calculated with the \emph{full} Hamiltonian of the system and the set of vectors $|\psi_{\mu}(\omega_{\mu})\ra$ can be normalized, while not necessarily being mutually orthogonal. The system of equations (\ref{eq:iteration_scheme_basis_zero_order_approximation_to_the_energy_of_the_system_3}), (\ref{eq:iteration_scheme_basis_zero_order_approximation_to_the_energy_of_the_system_4}) is the exact representation of the Schr\"odinger equation. For the approximate solution of this system, in accordance with OM the following concept is used: the closer the zeroth-order approximation of the state vector is to the exact solution, the closer the matrix $H_{\mu \nu}$ becomes to the diagonal one. Therefore, an iteration scheme for the solution of the system (\ref{eq:iteration_scheme_basis_zero_order_approximation_to_the_energy_of_the_system_3}), (\ref{eq:iteration_scheme_basis_zero_order_approximation_to_the_energy_of_the_system_4}) can be built, for which convergence is determined with the ratios of non-diagonal elements $H_{\mu \nu}$ to the diagonal ones $H_{\mu \mu}$ in the representation of the state vectors $|\psi_{\mu}(\omega_{\mu})\ra$. A sufficiently detailed discussion of the convergence of the iteration scheme for different physical systems is given in the monograph \cite{Feranchuk2015}. Consequently, we find the system of recurrent equations
\begin{align}
	E^{(s)}_{\mu} &= \left[1 + \sum_{\nu \neq \mu} C^{(s-1)}_{\mu \nu}I_{\mu \nu}\right]^{-1} \left[H_{\mu \mu} + \sum_{\nu \neq \mu} C^{(s-1)}_{\mu \nu}H_{\mu \nu}\right]; \label{eq:iteration_scheme_basis_zero_order_approximation_to_the_energy_of_the_system_5}
	\\
	C^{(s)}_{\mu \gamma} &= \left[E^{(s-1)}_{\mu} - H_{\gamma \gamma}\right]^{-1}\left[ H_{\gamma \mu} - E^{(s-1)}_{\mu}I_{\gamma \mu} + \sum_{\nu \neq \mu \neq \gamma}C^{(s-1)}_{\mu \nu} (H_{\gamma \nu} - E^{(s-1)}_{\mu}I_{\gamma \nu})\right]; \label{eq:iteration_scheme_basis_zero_order_approximation_to_the_energy_of_the_system_6}
	\\
	C^{(-1)}_{\mu \nu} &= C^{(0)}_{\mu \nu} = 0; \quad E^{(0)}_{\mu} = H_{\mu \mu}. \label{eq:iteration_scheme_basis_zero_order_approximation_to_the_energy_of_the_system_7}
\end{align}

As opposed to conventional perturbation theory, where the exact solution is defined as a sum of corrections of all orders, in OM the exact value of the energy of the system is given as a limit of a sequence
\begin{align}
	E_{\mu} = \lim_{s \rightarrow \infty} E^{(s)}_{\mu}; \quad s = 0,1,\ldots.\label{eq:iteration_scheme_basis_zero_order_approximation_to_the_energy_of_the_system_8}
\end{align}
In particular, for the first two iterations one can find
\begin{align}
	E^{(1)}_{\mu} &= E^{(0)}_{\mu} = H_{\mu \mu} ; \label{eq:iteration_scheme_basis_zero_order_approximation_to_the_energy_of_the_system_9}
	\\
	E^{(2)}_{\mu} &= \left[1 +   \sum_{\nu \neq \mu} \frac{(H_{\nu \mu} - E^{(0)}_{\mu}I_{\nu \mu})I_{\mu \nu}}{E^{(0)}_{\mu} - H_{\nu \nu} }\right]^{-1} \left[H_{\mu \mu} + \sum_{\nu \neq \mu} \frac{(H_{\nu \mu} - E^{(0)}_{\mu}I_{\nu \mu})H_{\mu \nu}}{E^{(0)}_{\mu} - H_{\nu \nu}}\right]. \label{eq:iteration_scheme_basis_zero_order_approximation_to_the_energy_of_the_system_10}
\end{align}

The last equation looks analogously to the second-order correction of perturbation theory, while the main difference is related to the denominators of equation (\ref{eq:iteration_scheme_basis_zero_order_approximation_to_the_energy_of_the_system_10}), where the matrix element $H_{\nu \nu}$ is calculated with the full Hamiltonian of the system, whereas the perturbation theory relations merely involve the diagonal element of the unperturbed Hamiltonian. As will be shown below, this is exactly the reason, which determines the convergence of integrals over intermediate states.

Before proceeding with the application of the iteration scheme, let us discuss the choice of the parameters $\{\omega_{\mu}\} = \{\omega_{\mu}^1,\ldots,\omega_{\mu}^n,\ldots\}$ in more detail. For this we note that the representation (\ref{eq:iteration_scheme_basis_zero_order_approximation_to_the_energy_of_the_system_3}-\ref{eq:iteration_scheme_basis_zero_order_approximation_to_the_energy_of_the_system_4}) is exactly equivalent to the Scr\"{o}dinger equation (\ref{eq:iteration_scheme_basis_zero_order_approximation_to_the_energy_of_the_system_1}), provided that the set of states $\{|\psi_{\mu}(\{\omega_{\mu}\})\ra\}$ is a complete one. It is evident that if the state vectors $\{|\psi_{\mu}(\{\omega_{\mu}\})\ra\}$ coincide with the exact eigenstates $|\Psi_{\mu}\ra$ of the full Hamiltonian, the matrix $H_{\mu\nu}$ is a diagonal one, i.e. $H_{\mu\nu} = E_{\mu}\delta_{\mu\nu}$ and the coefficients  $C_{\mu\nu} = 0$. The eigenvalues $E_{\mu}$ are determined exactly and are independent of the set of parameters $\{\omega_{\mu}\}$. Therefore, the relation
\begin{align}
	\frac{\partial E_{\mu}}{\partial\omega_{\mu}^n} \equiv 0, \quad n=\{1,2,\ldots\} \label{eq:iteration_scheme_basis_zero_order_approximation_to_the_energy_of_the_system_rev1}
\end{align}
holds identically.

According to our initial assumption we choose the trial parameters $\{\omega_{\mu}\}$ in the basis states $|\psi_{\mu}(\{\omega_{\mu}\})\ra$ such that they determine the best possible approximation for the exact solution $|\Psi_{\mu}\ra$ in the chosen class of functions. This is equivalent to the supposition that the off-diagonal elements of the matrix $H_{\mu \nu}$ are small numbers such that the ratios $H_{\mu \nu}/H_{\mu \mu}$ are proportional to some effective small parameter $\epsilon$. Therefore, the zeroth-order approximation of the operator method is chosen as
\begin{align}
	E_{\mu}^{(0)}(\{\omega_{\mu}\}) = H_{\mu \mu}(\{\omega_{\mu}\}), \quad C_{\mu \nu}^{(0)} = 0, \quad |\Psi_{\mu}^{(0)}\ra = |\psi_{\mu}(\{\omega_{\mu}\})\ra.\label{eq:iteration_scheme_basis_zero_order_approximation_to_the_energy_of_the_system_rev2}
\end{align}

However, the matrix $H_{\mu \nu}$ contains small off-diagonal elements, which need to be taken into account. Hence, the subsequent approximations read
\begin{align}
	E_{\mu} &= H_{\mu \mu}(\{\omega_{\mu}\}) + \sum_{s = 1}^{\infty}\epsilon^s E_{\mu}^{(s)}(\{\omega_{\mu}\}), \label{eq:iteration_scheme_basis_zero_order_approximation_to_the_energy_of_the_system_rev3}
	\\
	C_{\mu \nu}(\{\omega_{\mu}\}) &= \sum_{s=1}^{\infty}\epsilon^s C_{\mu \nu}^{(s)}(\{\omega_{\mu}\}),\quad \mu\neq \nu.\label{eq:iteration_scheme_basis_zero_order_approximation_to_the_energy_of_the_system_rev4}
\end{align}
As the left-hand side of equation (\ref{eq:iteration_scheme_basis_zero_order_approximation_to_the_energy_of_the_system_rev3}) does not depend on the parameters $\{\omega_{\mu}\}$ it is natural to require that in each order in $\epsilon$  the right-hand side also does not depend on $\{\omega_{\mu}\}$:
\begin{align}
	\frac{\partial E_{\mu}}{\partial\omega_{\mu}^n} = 0, \quad s=\{0,1,\ldots\}\label{eq:iteration_scheme_basis_zero_order_approximation_to_the_energy_of_the_system_rev5}
\end{align}
for each $\omega_{\mu}^n$. In the monograph \cite{Feranchuk2015} it was demonstrated that the recalculation of the parameters $\{\omega_{\mu}\}$ in every order in $\epsilon$ speeds up the convergence of the iteration scheme, however does not change the qualitative behaviour of the energy levels of the system. For this reason, in all calculations below we will fix the parameters $\{\omega_{\mu}\}$ via the zeroth-order approximation:
\begin{align}
	\frac{\partial E_{\mu}^{(0)}}{\partial\omega_{\mu}^n} = 0.\label{eq:iteration_scheme_basis_zero_order_approximation_to_the_energy_of_the_system_rev6}
\end{align}

In what follows we apply the above approach to the description of our model. In accordance with OM we choose a variational state vector, which incorporates the qualitative peculiarities of the system. From a physical point of view a field can be considered as a system of an infinite number of harmonic oscillators. Due to the interaction with a particle the equilibrium positions of these harmonic oscillators are modified. In a representation of creation and annihilation operators the shift of equilibrium positions corresponds to the displacement of a classical component $u_{\vec k}$ on these operators \cite{Scully1997}, i.e. $\opa a_{\vec k}^{\dag} \rightarrow \opa a_{\vec k}^{\dag}+u_{\vec k}^{*}$ and $\opa a_{\vec k} \rightarrow \opa a_{\vec k}+u_{\vec k}$, such that we choose a basis of field oscillators consisting of coherent states. As a result a so-called localized state of a particle in the field of these classical components arises. This means that during its existence the particle becomes ``dressed'', i.e. somewhat smeared out while still localized. Moreover, this ``dressed'' state should be an eigenstate of the total momentum operator $\opA P$, since $\opA P$ commutes with $\opa H$. Concluding, we formulate the following conditions for the state vectors: i) representation in the basis of coherent states; ii) imposing the localization of the particle state; iii) use of variational state vectors as eigenstates of $\opA P$ (\ref{eq:model_description4}).

In order to incorporate the first two conditions in the state vector, we choose it as the product of the square integrable wave function of a particle, localized near an arbitrary point $\vec R$ in space, and a coherent state of the field, analogous to the polaron problem \cite{RevModPhys.63.63,*Mitra198791,0022-3719-17-24-012}:
\begin{equation}
	\label{eq:iteration_scheme_basis_zero_order_approximation_to_the_energy_of_the_system_11}
	\vert\Psi({\vec r},{\vec R})\rangle= \phi({\vec r}-{\vec R})\exp\left(\sum_{\vec k}\Bigl(u^*_{\vec
k}e^{-\ri \thp{k}{R}} \opa a^\dag_{\vec k}-u_{\vec
k}e^{\ri \thp{k}{R}} \opa a_{\vec k}\Bigr)\right)\vert 0\rangle.
\end{equation}

In the state (\ref{eq:iteration_scheme_basis_zero_order_approximation_to_the_energy_of_the_system_11}) the classical component of the field $u_{\vec k}$ and the wave function $\phi(\vec r - \vec R)$ can be considered as the variational parameters $\{\omega_{\mu}\}$ of OM. In accordance with the above described procedure of the choice of the parameters $\{\omega_{\mu}\}$, the functional derivative over these parameters from the functional $\langle \Psi({\vec r},{\vec R})\vert \opa H  \vert\Psi({\vec r},{\vec R})\rangle$ should be equal to zero. This yields an equation for the classical components of the field $u_{\vec k}$ and the wave function $\phi(\vec r - \vec R)$:
\begin{align}
	\label{eq:iteration_scheme_basis_zero_order_approximation_to_the_energy_of_the_system_12}
	\frac{\delta}{\delta u_{\vec k}} &\left[ \langle \Psi({\vec r},{\vec R})\vert \opa H  \vert\Psi({\vec r},{\vec R})\rangle\right] = \frac{\delta}{\delta \phi({\vec r}-{\vec R})} \left[ \langle \Psi({\vec r},{\vec R})\vert \opa H  \vert\Psi({\vec r},{\vec R})\rangle\right] = 0.
\end{align}

By calculating the functional with the Hamiltonian (\ref{eq:model_description1}) and corresponding derivatives one obtains the connection between the classical components of the field and the wave function of the particle:
\begin{align}
	\label{eq:iteration_scheme_basis_zero_order_approximation_to_the_energy_of_the_system_13}
	u_{\vec k} &=-\frac{g}{\sqrt{2\Omega\omega^3_{\vec k}}}\int d \vec r |\phi ({\vec r})|^2 e^{-\ri\thp{k}{r}}.
\end{align}

In the general case, the second equation in (\ref{eq:iteration_scheme_basis_zero_order_approximation_to_the_energy_of_the_system_12}) leads to the integral equation for the function $\phi_{\vec P}(\vec r)$. However, according to reference \cite{Feranchuk2015}, the convergence of the iteration scheme of OM does not depend on the particular choice of variational parameters, under the condition that the approximate state vector takes into account qualitative characteristics of the system. Therefore, for the analytical investigation of the energy $E_L^{(0)}(\vec P,g)$ we replace the exact numerical solution with a trial wave function, which depends on the single parameter $\lambda$ and is equal to
\begin{align}
	\label{eq:iteration_scheme_basis_zero_order_approximation_to_the_energy_of_the_system_27}
	\phi(\vec r) = \frac{\lambda^{\frac{3}{2}}}{\pi^{\frac{3}{4}}}e^{-\frac{\lambda^2 r^2}{2}}.
\end{align}

We notice that in the polaron problem the application of OM with the trial wave function (\ref{eq:iteration_scheme_basis_zero_order_approximation_to_the_energy_of_the_system_27}) yields an accuracy of the order of $1\%$ in the calculation of the bound state energy and the effective mass \cite{0022-3719-17-24-012}. With this choice of wave function, we proceed to calculate the classical component of the field $u_{\vec k}$ and the Fourier transform of the wave function (\ref{eq:iteration_scheme_basis_zero_order_approximation_to_the_energy_of_the_system_27}), which will be required below:
\begin{align}
	u_{\vec k} &= -\frac{g}{\sqrt{2 \Omega}}\frac{1}{\sqrt{k^3}}\int d\vec r |\phi(\vec r)|^2 e^{-\ri\thp{k}{r}} = -\frac{g}{\sqrt{2 \Omega}}\frac{e^{-\frac{k^2}{4 \lambda^2}}}{\sqrt{k^3}}; \label{eq:iteration_scheme_basis_zero_order_approximation_to_the_energy_of_the_system_28}
	\\
	\phi_{\vec k} &= \int d\vec r \phi(\vec r)e^{-\ri\thp{k}{r}} = 2\sqrt{2}\frac{\pi^{\frac{3}{4}}}{\lambda^{\frac{3}{2}}} e^{-\frac{k^2}{2 \lambda^2}} = \phi_0 e^{-\frac{k^2}{2 \lambda^2}}. \label{eq:iteration_scheme_basis_zero_order_approximation_to_the_energy_of_the_system_29}
\end{align}

Furthermore, the states (\ref{eq:iteration_scheme_basis_zero_order_approximation_to_the_energy_of_the_system_11}) are not the eigenstates of the total momentum operator $\opA P$ of the system, i.e. they are not translationary invariant. Moreover, these states are degenerate, as they do not depend on the localization point $\vec R$ of the particle in space. The choice of the correct linear combination of these states allows one to build a set of states which are not degenerate and are eigenstates of the total momentum operator $\opA P$:
\begin{align}
	|\Psi^{(0)}_{\vec P_1,    n_{\vec k}}\ra &=  \frac{1}{N_{\vec P_1,n_{\vec k}}\sqrt{\Omega}}\int d \vec R \phi_{\vec P_1}(\vec r - \vec R)\exp\left\{\ri(\vec P_1 - \vec k n_{\vec k})\cdot \vec R \right\}\exp\left\{\sum_{\vec q}(u_{\vec q} e^{- \ri\vec q \cdot \vec R}\opa a_{\vec q}^\dag - u_{\vec q}^* e^{ \ri\vec q \cdot \vec R}\opa a_{\vec q})\right\}|n_{\vec k}\ra,\label{eq:iteration_scheme_basis_zero_order_approximation_to_the_energy_of_the_system_14}
	\\
	\opA P |\Psi^{(0)}_{\vec P_1,n_{\vec k}}\ra &= \vec P_1 |\Psi^{(0)}_{\vec P_1,n_{\vec k}}\ra. \nonumber
\end{align}
Here $\Omega$ is the normalization volume and $\vec P_1$ the total momentum of the system, $|n_{\vec k}\ra$ are Fock field states with occupation number $n_{\vec k}$, $\phi_{\vec P_1}(\vec r - \vec R)$ is the wave function of the particle localized at point $\vec r = \vec R$ and the classical component of the field $u_{\vec k}$ is defined via equation (\ref{eq:iteration_scheme_basis_zero_order_approximation_to_the_energy_of_the_system_13}). The normalization constant for the state $|\Psi^{(0)}_{\vec P_1, 1_{\vec k}}\ra$ is defined as
\begin{align}
	\label{eq:iteration_scheme_basis_zero_order_approximation_to_the_energy_of_the_system_15}
	|N_{\vec P_1,1_{\vec k}}|^2 = \int d \vec R_1 d \vec \rho \phi_{\vec P_1}^*(\vec \rho)\phi_{\vec P_1}(\vec \rho - \vec R_1) e^{\ri(\vec P_1 - \vec k)\cdot \vec R_1 + \sum_k |u_k|^2(e^{-\ri\thp{k}{R_1}}-1)}\left(2|u_k|^2(\cos\thp{k}{R_1}-1)+1\right).
\end{align}

In addition the set of states (\ref{eq:iteration_scheme_basis_zero_order_approximation_to_the_energy_of_the_system_14}) forms a complete and an orthonormal basis in Hilbert space. The completeness of these states follows from the fact that they are eigenstates of a Hermitian operator $\opA{P}$, with an explicit proof given in Appendix~A. Concluding, the set of states (\ref{eq:iteration_scheme_basis_zero_order_approximation_to_the_energy_of_the_system_14}) takes into account physical peculiarities of the system, forms the complete set of states in Hilbert space for arbitrary functions $\phi_{\vec P}(\vec r)$ and $u_{\vec k}$, which are an analog to the parameters $\{\omega_{\mu}\}$, and, therefore, can be usable in the iteration scheme (\ref{eq:iteration_scheme_basis_zero_order_approximation_to_the_energy_of_the_system_5}-\ref{eq:iteration_scheme_basis_zero_order_approximation_to_the_energy_of_the_system_7}).

The zeroth-order approximation for the ground state vector following above procedure then reads as
\begin{align}
	\label{eq:iteration_scheme_basis_zero_order_approximation_to_the_energy_of_the_system_16}
	|\Psi^{(L)}_{\vec P}\ra   &= \frac{1}{N_{\vec P}\sqrt{\Omega}} \int d{\vec R}\,\phi_{\vec P}({\vec r}-{\vec R})\exp \left(\ri\thp{P}{R}+   \sum_{\vec k}\Bigl(u^*_{\vec k}e^{-\ri \thp{k}{R}} \opa a^\dag_{\vec k}-u_{\vec k}e^{\ri \thp{k}{R}} \opa a_{\vec k}\Bigr)\right)| 0 \ra,
\end{align}
whereas equations (\ref{eq:iteration_scheme_basis_zero_order_approximation_to_the_energy_of_the_system_9}), (\ref{eq:iteration_scheme_basis_zero_order_approximation_to_the_energy_of_the_system_10}) look like
\begin{align}
	E^{(2)}  &= \frac{E_L^{(0)} +   \sum_{\vec P_1, \{n_k \neq 0\}}C^{(1)}_{\vec P_1, \{n_{\vec k}\}}\la\Psi^{(L)}_{\vec P}| \opa H |\Psi_{\vec P_1,   \{ n_{\vec k}\}}\ra}{1 + \sum_{\vec P_1, \{n_k\neq 0\}}C^{(1)}_{\vec P_1, \{n_{\vec k}\}}\la\Psi^{(L)}_{\vec P}| \Psi_{\vec P_1,   \{ n_{\vec k}\}}\ra};\label{eq:iteration_scheme_basis_zero_order_approximation_to_the_energy_of_the_system_17}
	\\
C^{(1)}_{\vec P_1, \{n_{\vec k}\}}  &= \frac{E_L^{(0)} \la \Psi_{\vec P_1,   \{ n_{\vec k}\}} |\Psi^{(L)}_{\vec P}\ra   - \la \Psi_{\vec P_1,   \{ n_{\vec k}\}}| \opa H |\Psi^{(L)}_{\vec P}\ra }{H_{\vec P_1, \{n_{\vec k}\}; \vec P_1, \{n_{\vec k}\}} - E_L^{(0)}}; \label{eq:iteration_scheme_basis_zero_order_approximation_to_the_energy_of_the_system_18}
	\\
H_{\vec P_1, \{n_{\vec k}\}; \vec P_2, \{n_{1\vec k}\}} &= \la \Psi_{\vec P_1,   \{ n_{\vec k}\}}| \opa H |\Psi_{\vec P_2,   \{ n_{1\vec k}\}}\ra, \quad E_L^{(0)} = \la \Psi^{(L)}_{\vec P}| \opa H |\Psi^{(L)}_{\vec P}\ra. \nonumber
\end{align}

We want to emphasize once more that all matrix elements are calculated with the full Hamiltonian of the system
\begin{align}
	\label{eq:iteration_scheme_basis_zero_order_approximation_to_the_energy_of_the_system_19}
	\opa H = \frac{1}{2}\left( \vec{P}^2 - 2 \sum_{\vec k} \opa a_{\vec k}^\dag \opa a_{\vec k} \vec k \cdot  \vec{P} +\left(\sum_{\vec k} \opa a_{\vec k}^\dag \opa a_{\vec k} \vec k\right)^2\right)+\sum_{\vec k}\omega_{\vec k} \opa a^\dag_{\vec k}\opa a_{\vec k}+ \frac{g}{\sqrt{\Omega}}\sum_{\vec k}\frac{1}{\sqrt{2\omega_{\vec k}}} \left(e^{\ri\vec k\vec r}\opa a_{\vec k}+e^{-\ri\vec k\vec r}\opa a^\dag_{\vec k}\right).
\end{align}

Let us calculate the ground state energy $E^{(0)}_L$ of the system in this basis. The details of the calculations can be found in appendix C. The ground state energy reads accordingly
\begin{align}
	&E_{L}^{(0)}(\vec P,g)= \frac{P^2}{2} - \thp{P}{Q} + G + E_{\text{f}}(\vec P) + E_{\text{int}}(\vec P), \label{eq:iteration_scheme_basis_zero_order_approximation_to_the_energy_of_the_system_20}
\end{align}
with
\begin{align*}
	\vec Q &= \frac{1}{|N_{\vec P}|^2} \sum_{\vec k}\vec k |u_{\vec k}|^2\int d{\vec  R} d{\vec r}\,\phi_{\vec P}^*({\vec r})\phi_{\vec P}({\vec r - \vec R}) e^{\Phi (\vec R) + \ri (\vec P - \vec k)\cdot \vec R}; 
	\\
	 G &=\frac{1}{2} \frac{1}{|N_{\vec P}|^2}\sum_{\vec m,\vec l}\thp{m}{l}|u_{\vec m}|^2 |u_{\vec l}|^2\int d\vec r d\vec R \phi^*(\vec r)\phi(\vec r - \vec R)e^{\ri\thp{P}{R}+\Phi(\vec R)-\ri(\vec m + \vec l)\cdot R};
	\\
	E_{\text{f}}(\vec P) &= \frac{1}{|N_{\vec P}|^2} \sum_{\vec k}\left(k + \frac{k^2}{2}\right)|u_{\vec k}|^2\int d{\vec  R} d{\vec r}\,\phi_{\vec P}^*({\vec r})\phi_{\vec P}({\vec r-\vec R}) e^{\Phi (\vec R) + \ri (\vec P - \vec k)\cdot \vec R};
	\\
	E_{\text{int}}(\vec P) &= \frac{g}{|N_{\vec P}|^2 } \sum_{\vec k}\frac{u_{\vec k}}{\sqrt{2 k \Omega}} \int d{\vec  R} d{\vec r}\left(\phi_{\vec P}^*({\vec r}+\vec R)\phi_{\vec P}({\vec r})+\phi_{\vec P}^*({\vec r})\phi_{\vec P}({\vec r}-{\vec R})\right)e^{\Phi (\vec R) + \ri (\thp{P}{R} + \thp{k}{r})};
	\\
	\Phi (\vec R)&= \sum_{\vec k}|u_{\vec k}|^2(e^{-\ri\thp{k}{R}}-1);
	\\
	|N_{\vec P}|^2 &= \int d{\vec  R} d{\vec r}\,\phi_{\vec P}^*({\vec r})\phi_{\vec P}({\vec r}-{\vec R}) e^{\Phi(\vec R) + \ri \thp{P}{R}}.
\end{align*}

Actually, the iteration scheme (\ref{eq:iteration_scheme_basis_zero_order_approximation_to_the_energy_of_the_system_5}), (\ref{eq:iteration_scheme_basis_zero_order_approximation_to_the_energy_of_the_system_6}), (\ref{eq:iteration_scheme_basis_zero_order_approximation_to_the_energy_of_the_system_7}) can be used for arbitrary coupling constants \cite{Feranchuk2015}. However, as was described above, in the framework of our model we are interested in the behavior of the ground state energy $E_L^{(0)}$ in the weak coupling limit. In this limit we can neglect the function 
\begin{align*}
	\Phi(\vec R) = \sum_{\vec m}|u_{\vec m}|^2 \left(e^{-\ri\thp{m}{R}}-1\right)\sim g^2,
\end{align*}
in the exponent of all integrals in equation (\ref{eq:iteration_scheme_basis_zero_order_approximation_to_the_energy_of_the_system_20}) as $g\ll1$.

First of all, we investigate the situation of a particle at rest, i.e. $\vec P = 0$. In this case for the weak coupling limit the integrals in equation (\ref{eq:iteration_scheme_basis_zero_order_approximation_to_the_energy_of_the_system_20}) can be expressed through the Fourier transforms of the wave function $\phi(\vec r)$:
\begin{align}
	\int d\vec R_1 d \vec \rho \phi^*(\vec \rho)\phi(\vec \rho - \vec R_1) e^{-\ri\thp{k}{R_1}} =\int d \vec \rho \phi^*(\vec \rho)e^{-\ri\thp{k}{\rho}}\int d \vec R \phi(\vec R)e^{\ri\thp{k}{R}} = \phi^*_{\vec k}\phi_{-\vec k} &= \phi_{\vec k}^2, \label{eq:iteration_scheme_basis_zero_order_approximation_to_the_energy_of_the_system_30}
	\\
	\int d\vec R_1 d \vec \rho \phi^*(\vec \rho)\phi(\vec \rho - \vec R_1) = |\phi_0|^2 &= \phi_0^2, \label{eq:iteration_scheme_basis_zero_order_approximation_to_the_energy_of_the_system_31}
	\\
	\int d \vec R_1 d \vec \rho \phi^*(\vec \rho)\phi(\vec \rho - \vec R_1) e^{-\ri\thp{k}{\rho}} = \phi^*_{\vec k}\phi_0 &= \phi_{\vec k}\phi_0. \label{eq:iteration_scheme_basis_zero_order_approximation_to_the_energy_of_the_system_32}
\end{align}

With the use of equations (\ref{eq:iteration_scheme_basis_zero_order_approximation_to_the_energy_of_the_system_20}), (\ref{eq:iteration_scheme_basis_zero_order_approximation_to_the_energy_of_the_system_30}-\ref{eq:iteration_scheme_basis_zero_order_approximation_to_the_energy_of_the_system_32}) we can rewrite the energy of the ground state in a form
\begin{align}
	E^{(0)}_L(0, g) = \frac{1}{2}\sum_{\vec m,\vec l}\thp{m}{l}|u_{\vec m}|^2 |u_{\vec l}|^2\,\frac{\phi_{\vec l+ \vec m}^2}{\phi_0^2}+\sum_{\vec k}\left(k + \frac{k^2}{2}\right)|u_{\vec k}|^2 \frac{\phi_{\vec k}^2}{\phi_0^2} + \frac{2g}{\sqrt{2 \Omega}}\sum_{\vec k}\frac{u_{\vec k}}{\sqrt{k}}\frac{\phi_{\vec k}}{\phi_0}, \label{eq:iteration_scheme_basis_zero_order_approximation_to_the_energy_of_the_system_33}
\end{align}
which up to fourth order in $g$ yields
\begin{align}
	E^{(0)}_L(0, g)  =  \frac{g^2}{24\pi^2}\left(\lambda(-4+\sqrt{2})\sqrt{3\pi}+\lambda^2\right) + O (g^4).\label{eq:iteration_scheme_basis_zero_order_approximation_to_the_energy_of_the_system_34}
\end{align}

By minimizing the energy with respect to $\lambda$ one finds
\begin{align}
	E^{(0)}_L(0, g) = - g^2\frac{(-4+\sqrt{2})^2}{32\pi}; \quad \lambda = \frac{\sqrt{3\pi}}{2}(4 - \sqrt{2}).\label{eq:iteration_scheme_basis_zero_order_approximation_to_the_energy_of_the_system_35}
\end{align}

In the weak-coupling limit it is also possible to obtain a renormalization for the mass of the particle. This is accomplished by expanding the energy (\ref{eq:iteration_scheme_basis_zero_order_approximation_to_the_energy_of_the_system_20}) in a series over $\vec P$ up to second order. The details of the calculation can be found in appendix D. The result reads
\begin{align}
	E^{(0)}_L(P, g)\approx  E^{(0)}_L(0, g) + \frac{P^2}{2}\left[1-\frac{g^2}{9\pi^2}\frac{17-\sqrt{2}}{21}\right]; \quad m^{(0)*} = 1+\frac{g^2}{9\pi^2}\frac{17-\sqrt{2}}{21}. \label{eq:iteration_scheme_basis_zero_order_approximation_to_the_energy_of_the_system_36}
\end{align}

From this equation we can conclude that the factor, which determines the corrected mass is half the one via the leading second-order term from perturbation theory, see e.g. equation (\ref{eq:model_description14}). 

\section{Second order iteration for the energy and convergence} 
\label{sec:second_order_iteration_for_the_energy_convergence}
In the previous section we have found the energy of the ground state and the renormalized mass, which are proportional to the square of the coupling constant in zeroth-order approximation. However, the correction to the energy coming from single-phonon intermediate transitions is of the same order with respect to the coupling constant. Consequently, its contribution should also be taken into account, thus requiring the calculation of the energy of the system in the second iteration (\ref{eq:iteration_scheme_basis_zero_order_approximation_to_the_energy_of_the_system_17}).

In order to calculate the second order iteration for the energy we notice (appendix E) that the matrix elements $\la \Psi_{\vec P_1,   \{ n_{\vec k}\}} |\Psi^{(L)}_{\vec P}\ra$ and $\la \Psi_{\vec P_1,   \{ n_{\vec k}\}}| \opa H |\Psi^{(L)}_{\vec P}\ra$, which are found in equations (\ref{eq:iteration_scheme_basis_zero_order_approximation_to_the_energy_of_the_system_17}), (\ref{eq:iteration_scheme_basis_zero_order_approximation_to_the_energy_of_the_system_18}) are proportional to the delta function of the total momentum of the system $\delta(\vec P_1 - \vec P)$. Therefore, during the evaluation of the sum over $\vec P_1$ in equation (\ref{eq:iteration_scheme_basis_zero_order_approximation_to_the_energy_of_the_system_17}) for the energy we have used the usual procedure \cite{LandauQED}: one of the delta functions in its square was replaced through the normalization volume $\Omega$, and the integration over the remaining one yields $\vec P = \vec P_1$, thus expressing the conservation of momentum.

Firstly, we consider the case, when a particle is at rest, i. e. $\vec P = 0$. The results, which are expressed through the Fourier components of the particle wave function in the weak coupling limit read:
\begin{align}
	 E^{(2)}(0,g) = \frac{A}{B},\label{eq:second_order_iteration_for_the_energy_convergence_1}
\end{align}
where
\begin{align}
	A = E_L^{(0)}&+\sum_{\vec k}\frac{1}{\phi_{\vec k}^2 \phi_0^2}\left[-u_{\vec k} \phi_{\vec k}^2\left(\frac{k^2}{2}+k\right)-\frac{g}{\sqrt{2 \Omega}}\frac{\phi_{\vec k} \phi_0}{\sqrt{k}}-u_{\vec k} \phi_{\vec k}^2\bigg(g^2 I_{\vec k}+g^4 J_{\vec k}-E_L^{(0)}\bigg) \right] \nonumber
	\\
	&\mspace{40mu}\times\left[u_{\vec k} \phi_{\vec k}^2\left(\frac{k^2}{2}+k\right)+\frac{g}{\sqrt{2 \Omega}}\frac{\phi_{\vec k} \phi_0}{\sqrt{k}}+u_{\vec k} \phi_{\vec k}^2\bigg(g^2 I_{\vec k}+g^4 J_{\vec k}\bigg) - E_L^{(0)} u_{\vec k} \phi_0^2\right] \nonumber
	\\
	&\mspace{40mu}\times \left[\left(\frac{k^2}{2}+k\right)+g^2 I_{\vec k}+g^4 J_{\vec k}-E_L^{(0)}\right]^{(-1)}, \label{eq:second_order_iteration_for_the_energy_convergence_2}
\end{align}
and
\begin{align}
	B = 1 + \sum_{\vec k}\frac{1}{\phi_{\vec k}^2 \phi_0^2}\frac{\left[-u_{\vec k} \phi_{\vec k}^2\left(\frac{k^2}{2}+k\right)-\frac{g}{\sqrt{2 \Omega}}\frac{\phi_{\vec k} \phi_0}{\sqrt{k}}-u_{\vec k} \phi_{\vec k}^2\bigg(g^2 I_{\vec k}+g^4 J_{\vec k}-E_L^{(0)}\bigg) \right]u_{\vec k}(\phi_{\vec k}^2-\phi_{0}^2)}{\left(\frac{k^2}{2}+k\right)+g^2 I_{\vec k}+g^4 J_{\vec k}-E_L^{(0)}}.\label{eq:second_order_iteration_for_the_energy_convergence_3}
\end{align}

In equations (\ref{eq:second_order_iteration_for_the_energy_convergence_2}-
\ref{eq:second_order_iteration_for_the_energy_convergence_3}) we have introduced the following notations
\begin{align}
\sum_{\vec m}\vec m |u_{\vec m}|^2 \phi_{\vec m+\vec k}^2 &\equiv	g^2 \phi_{\vec k}^2 \vec I^{(1)}_{\vec k}; \quad
\vec I^{(1)}_{\vec k}  =
	  \frac{\vec k}{k^2}\frac{\lambda^2}{32\pi^2} \frac{4k-e^{\frac{2}{3}\frac{k^2}{\lambda^2}}\sqrt{6\pi}\lambda \text{Erf}\frac{\sqrt{\frac{2}{3}}k}{\lambda}}{k}; \label{eq:second_order_iteration_for_the_energy_convergence_4}
	  \\
	  \sum_{\vec m}\left(\frac{m^2}{2}+m\right)|u_{\vec m}|^2 \phi_{\vec m+\vec k}^2 &\equiv	g^2 \phi_{\vec k}^2 I^{(2)}_{\vec k}; \quad I^{(2)}_{\vec k} =  \frac{ \lambda^2}{96 \pi^2} \frac{\sqrt{6\pi}\lambda e^{\frac{2}{3}\frac{k^2}{\lambda^2}}\text{Erf}\frac{\sqrt{\frac{2}{3}}k}{\lambda} + 6 \pi \text{Erfi}\frac{\sqrt{\frac{2}{3}}k}{\lambda}}{k}; \label{eq:second_order_iteration_for_the_energy_convergence_5}
	   \\
	   \frac{g \phi_{\vec k}}{\sqrt{2 \Omega}}\sum_{\vec m}\frac{u_{\vec m}}{\sqrt{m}}(\phi_{\vec m+\vec k}+\phi_{\vec m - \vec k}) &\equiv	g^2 \phi_{\vec k}^2 I^{(3)}_{\vec k}; \quad   I^{(3)}_{\vec k} =    - \frac{  \lambda^2}{4\pi}\frac{\text{Erfi}\frac{k}{\sqrt{3}\lambda}}{k}; \label{eq:second_order_iteration_for_the_energy_convergence_6}
	   \\
	   I_{\vec k} &= \vec k \cdot \vec I_{\vec k}^{(1)}+I_{\vec k}^{(2)}+I_{\vec k}^{(3)}; \label{eq:second_order_iteration_for_the_energy_convergence_7}
	   \\
	   \frac{1}{2}\sum_{\vec l,\vec m}\thp{l}{m}|u_{\vec l}|^2|u_{\vec m}|^2 \phi_{\vec l+\vec m+\vec k}^2 &\equiv g^4 \phi_{\vec k}^2 J_{\vec k}; \quad J_{\vec k} \approx \frac{5^{1/2}\lambda^2}{4(2\pi)^3 3^5}e^{\frac{4}{5}\frac{k^2}{\lambda^2}}\frac{\frac{2}{15}\frac{k^2}{\lambda^2} - 1}{(1+\frac{4}{45}\frac{k^2}{\lambda^2})^3}, \label{eq:second_order_iteration_for_the_energy_convergence_8}
\end{align}
where $\text{Erf}(x) = 2/\sqrt{\pi}\int_0^x e^{-z^2}dz$ and $\text{Erfi}(x) = -\ri\text{Erf}(\ri x)$ are the error function and the imaginary error functions, respectively. When we calculated the energy of the ground state, we dropped all terms with power in $g$ larger than $g^2$. Consequently, we can neglect the term $g^4 J_{\vec k}$ in comparison with $g^2 I_{k}$, which can be confirmed by the direct numerical calculation of the integral.

Prior to the numerical evaluation of the integrals (\ref{eq:second_order_iteration_for_the_energy_convergence_2}) and (\ref{eq:second_order_iteration_for_the_energy_convergence_3}), let us understand their structure through the approximate analytical calculation. We investigate the behavior of the numerator and denominator of the quantities $A$ and $B$. We start from breaking the integration region into two parts, namely $[0,k_0]$ and $[k_0,\infty)$. The value $k_0$ will be fixed below. Let us work out the behavior of the quantity $I_{\vec k}$ for small and large values of $k$. First of all we notice that $g^2 I_{0}$ gives exactly the ground state energy $E_L^{(0)}$. For small values of $k$, with the increase of $k$ the value of $g^2 I_{\vec k}\sim -g^2 k^2/(18\pi^2) +E_L^{(0)}$, i.e, it grows quadratically in absolute value, while being negative. Therefore, due to the presence of $g^2$, this term is small in comparison with $k^2/2+k$ for small values of $k$, so that, in the denominator of quantity $A$, the leading term is $k^2/2+k$. For large values of $k$, the quantity $g^2 I_{\vec k}$   exponentially grows as $I_{\vec k}\sim e^{\frac{2}{3}\frac{k^2}{\lambda^2}}/k$ and becomes the leading contribution in comparison with $k^2/2+k$, despite the higher power of $g$.

In analogy, we can analyze the numerator of the quantity $A$. For small values of $k$ we can neglect in every square bracket in equation (\ref{eq:second_order_iteration_for_the_energy_convergence_2}) the large powers of $g$, i.e. terms with exponents larger than $1$. Consequently, for small values of $k$, the integrand within $A$ looks like
\begin{align}
	\label{eq:second_order_iteration_for_the_energy_convergence_9}
	-\frac{\left[u_{\vec k} \phi_{\vec k}^2\left(\frac{k^2}{2}+k\right)+\frac{g}{\sqrt{2 \Omega}}\frac{\phi_{\vec k} \phi_0}{\sqrt{k}}\right]^2}{\phi_{\vec k}^2 \phi_0^2 \left(\frac{k^2}{2}+k\right)}.
\end{align}

For large values of $k$, the numerator is exponentially decreasing, with the leading term being $(-\frac{g}{\sqrt{2 \Omega}}\frac{\phi_{\vec k} \phi_0}{\sqrt{k}})(-E_L^{(0)} u_{\vec k})$. This follows from the fact that $u_{\vec k}\sim e^{-\frac{k^2}{4 \lambda^2}}$ and $\phi_{\vec k}\sim e^{-\frac{k^2}{2 \lambda^2}}$. Consequently, the integrand for large values of $k$ can be presented as
\begin{align}
	\label{eq:second_order_iteration_for_the_energy_convergence_10}
	\frac{(-\frac{g}{\sqrt{2 \Omega}}\frac{\phi_{\vec k} \phi_0}{\sqrt{k}})(-E_L^{(0)} u_{\vec k})}{g^2\phi_{\vec k}^2 I_{\vec k}}.
\end{align}

Combining all together, we find that the quantity $A$ can be approximately calculated as
\begin{align}
	A \approx E_L^{(0)} + \sum_{\vec k<\vec k_0}\frac{-\left(u_{\vec k} \frac{\phi_{\vec k}}{\phi_0}\left(\frac{k^2}{2}+k\right)+\frac{g}{\sqrt{2 \Omega}}\frac{1}{\sqrt{k}}\right)^2}{(\frac{k^2}{2}+k)}+\sum_{\vec k>\vec k_0}\frac{(-\frac{g}{\sqrt{2 \Omega}}\frac{\phi_{\vec k} \phi_0}{\sqrt{k}})(-E_L^{(0)} u_{\vec k})}{\phi_{\vec k}^2 g^2 I_{\vec k}}. \label{eq:second_order_iteration_for_the_energy_convergence_11}
\end{align}
In this expression, both sums are well defined and remain finite. The sum over the region $k>k_0$ is finite and convergent, while the ratio of numerator and denominator in the integrand is exponentially decreasing as $e^{-\frac{5k^2}{12 \lambda^2}}$.

In equation (\ref{eq:second_order_iteration_for_the_energy_convergence_11}) the point $k_0$ is determined as a solution of the equation
\begin{align}
	\frac{k^2}{2} + k + g^2 I_{\vec k} - E_L^{(0)} = 0, \label{eq:second_order_iteration_for_the_energy_convergence_12}
\end{align}
or employing the asymptotic behavior for the function $I_{\vec k}$ (appendix F)
\begin{align}
	\frac{k_0^2}{2} + k_0 = g^2\frac{\lambda^3 \sqrt{6\pi}}{48\pi^2}\frac{e^{\frac{2}{3}\frac{k_0^2}{\lambda^2}}}{k_0},
\label{eq:second_order_iteration_for_the_energy_convergence_13}
\end{align}
and by finding the following logarithm
\begin{align}
	\ln\frac{(\frac{k_0^2}{2}+k_0)k_0}{a} &= -2|\ln g|+\frac{2}{3}\frac{k_0^2}{\lambda^2},\label{eq:second_order_iteration_for_the_energy_convergence_14}
\end{align}
with
\begin{align*}
	a &= \frac{\lambda^3 \sqrt{6\pi}}{48\pi^2}.
\end{align*}

In the limit of extremely small $g$, we can build the solution of equation (\ref{eq:second_order_iteration_for_the_energy_convergence_14}) via iterations, thus yielding
\begin{align}
	k_0 \sim \lambda\sqrt{3|\ln g|}.\label{eq:second_order_iteration_for_the_energy_convergence_15}
\end{align}
The estimation of quantity $B$ can be performed in a similar fashion and one finds
\begin{align}
	B \approx 1+ \sum_{\vec k<\vec k_0}\frac{-\left(u_{\vec k} \frac{\phi_{\vec k}^2}{\phi_0^2}\left(\frac{k^2}{2}+k\right)+\frac{g}{\sqrt{2 \Omega}}\frac{\phi_{\vec k}}{\phi_0\sqrt{k}}\right)u_{\vec k}(\phi_{\vec k}^2-\phi_0^2)}{\phi_{\vec k}^2(\frac{k^2}{2}+k)}+ \sum_{\vec k>\vec k_0}\frac{(-\frac{g}{\sqrt{2 \Omega}}\frac{\phi_{\vec k} \phi_0}{\sqrt{k}})u_{\vec k}(\frac{\phi_{\vec k}^2}{\phi_0^2}-1)}{\phi_{\vec k}^2 g^2 I_{\vec k}}.\label{eq:second_order_iteration_for_the_energy_convergence_16}
\end{align}

At first sight, it may appear that the quantity $B$ features an infrared divergence, because the term $\frac{g}{\sqrt{2 \Omega}}\frac{\phi_{\vec k} u_{\vec k}}{\phi_0\sqrt{k}}/(\phi_{\vec k}^2(\frac{k^2}{2}+k)) \sim 1/k^3$ as $k\rightarrow 0$. However, this additional power of $k$ in the denominator is cancelled through the difference $\phi_{\vec k}^2 - \phi_0^2 \sim k^2/\lambda^2$. The convergence at infinity is manifested with the exponential decrease of the integrand $\sim e^{-\frac{5k^2}{12 \lambda^2}}$.

By plugging into equations (\ref{eq:second_order_iteration_for_the_energy_convergence_11}) and (\ref{eq:second_order_iteration_for_the_energy_convergence_16}) the values of $\phi_{\vec k}$ and $u_{\vec k}$, which are defined in equations (\ref{eq:iteration_scheme_basis_zero_order_approximation_to_the_energy_of_the_system_28}-
\ref{eq:iteration_scheme_basis_zero_order_approximation_to_the_energy_of_the_system_29}) and calculating the integrals (appendix F) we find the approximate analytical formula for the second iteration for the ground state energy
\begin{align}
	A &\approx E_L^{(0)}-\left[\frac{g^2\lambda}{24 \pi ^2}\left(\sqrt{6 \pi } \text{Erf}\left(\frac{\sqrt{\frac{3}{2}} k_0}{\lambda }\right)+\lambda -\lambda e^{-\frac{3k_0^2}{2 \lambda ^2}}\right)-\frac{g^2\lambda}{2 \sqrt{3}\pi^{3/2}}\text{Erf}\left(\frac{\sqrt{3}k_0}{2\lambda }\right)\right]-\frac{g^2}{2\pi^2}\ln\left(\frac{k_0}{2}+1\right)+E_L^{(0)}\frac{12\sqrt{6\pi}}{5 \lambda \pi}e^{-\frac{5k_0^2}{12 \lambda^2}}, \label{eq:second_order_iteration_for_the_energy_convergence_17}
	\\
	B &\approx 1+\frac{g^2}{12\pi^2}(1-e^{-\frac{3}{2}\frac{k_0^2}{\lambda^2}})-g^2f\left(\frac{k_0}{\lambda}\right) - \frac{144\sqrt{6\pi}}{25 \lambda \pi}\left(1+\frac{5}{12}\frac{k_0^2}{\lambda^2}\right)e^{-\frac{5k_0^2}{12 \lambda^2}}, \label{eq:second_order_iteration_for_the_energy_convergence_18}
	\\
	f(x) &= \frac{1}{4\pi^2}\int_0^x \frac{tdt}{1+t/2}e^{-\frac{3}{4}t^2}.\nonumber
\end{align}
Within the accuracy of the approximate formulas, we can set $B\approx1$. Therefore, one finally obtains
\begin{align}
	E^{(2)}(0,g) \approx A. \label{eq:second_order_iteration_for_the_energy_convergence_19}
\end{align}

The use of our simple analytical expressions allows to establish the behavior of the energy as a function of the coupling constant and consequently to determine the character of the singularity. In order to select the singularity, we investigate the limit
\begin{align*}
	\lim_{g\rightarrow0}E^{(2)}(0,g).
\end{align*}
In this limit, the value of $k_0$ logarithmically grows. Consequently, we can approximately set $k_0 \rightarrow \infty$ both in the expression in square brackets and in the last term of equation (\ref{eq:second_order_iteration_for_the_energy_convergence_17}). This way, the square bracket becomes equal to the energy of the ground state (appendix F) and cancels $E_L^{(0)}$. The last term also does not contribute to the energy as being exponentially small. Consequently, the only term remains, which exactly determines the character of the singularity and is equal to
\begin{align}
	E^{(2)}(0,g) &\underset{g\rightarrow0}{\longrightarrow}-\frac{g^2}{2\pi^2}\ln\left(\frac{k_0}{2}+1\right);
                                                 \label{eq:second_order_iteration_for_the_energy_convergence_20}
	\\
	k_0 &\approx \lambda\sqrt{3|\ln g|}. \label{eq:second_order_iteration_for_the_energy_convergence_21}
\end{align}
We observe that this term exactly coincides with the result via perturbation theory, i.e. equation (\ref{eq:model_description13}), however, here with a well specified ``cut-off''. Moreover, the most contributions to the integral in the energy arise from the region $k<k_0$ and this is exactly the reason for the natural ``cut-off'', which is determined self consistently and is directly related to the only parameter of the Hamiltonian, namely the coupling constant. Let us mention here that the corrections to the energy of the system (\ref{eq:second_order_iteration_for_the_energy_convergence_20}) arise in the subsequent iteration and are related to the transitions into intermediate states with two phonons. These contributions are proportional to $g^4$.

In addition we note here that the absence of the ultraviolet divergence in the energy of the ground state in equations (\ref{eq:second_order_iteration_for_the_energy_convergence_1}, \ref{eq:second_order_iteration_for_the_energy_convergence_2}, \ref{eq:second_order_iteration_for_the_energy_convergence_3}) is due to the fact that as in the zeroth-order approximation and in the second-order iteration in the resolvent $[E^{(0)}_{\mu} - H_{\nu \nu}]^{-1}$ of equation (\ref{eq:iteration_scheme_basis_zero_order_approximation_to_the_energy_of_the_system_10}) the dressed wave functions (\ref{eq:iteration_scheme_basis_zero_order_approximation_to_the_energy_of_the_system_14}) were used. This leads to the effective momentum cut-off $k_0(g)$, which is determined as the solution of equation (\ref{eq:second_order_iteration_for_the_energy_convergence_12}). This cut-off is a function of the coupling constant and is not a phenomenological parameter, which needs to be introduced for the removal of the ultraviolet divergence. Moreover, as follows from equation (\ref{eq:second_order_iteration_for_the_energy_convergence_20}), the energy of the ground state has a logarithmic singularity as $g\rightarrow 0$. It is clearly seen that this dependence can not be sorted out in the framework of perturbation theory, which yields a power series over the coupling constant $g$.

\begin{figure}[t]
	\centering
		\includegraphics[width=0.44\textwidth]{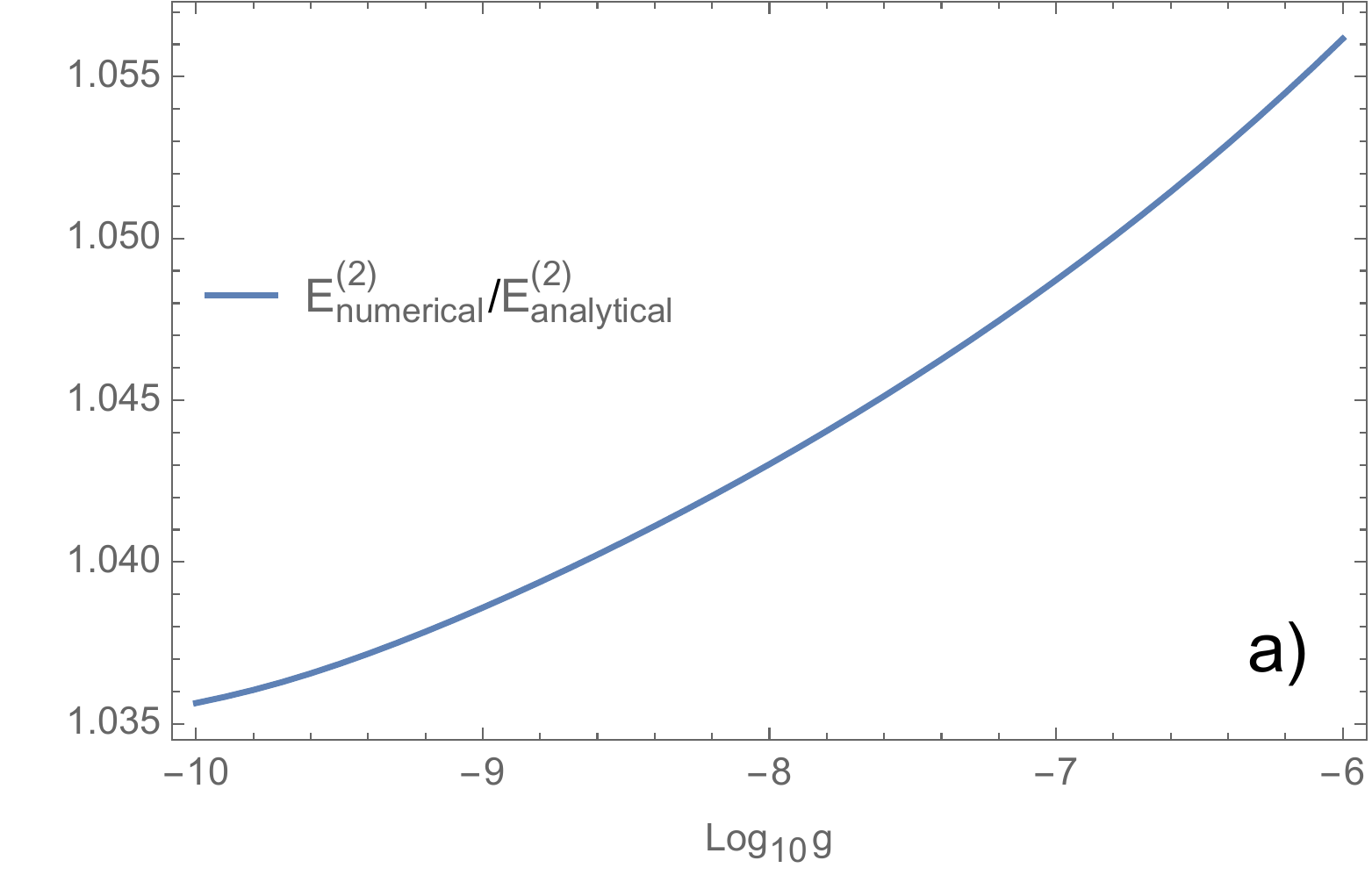}
		\includegraphics[width=0.48\textwidth]{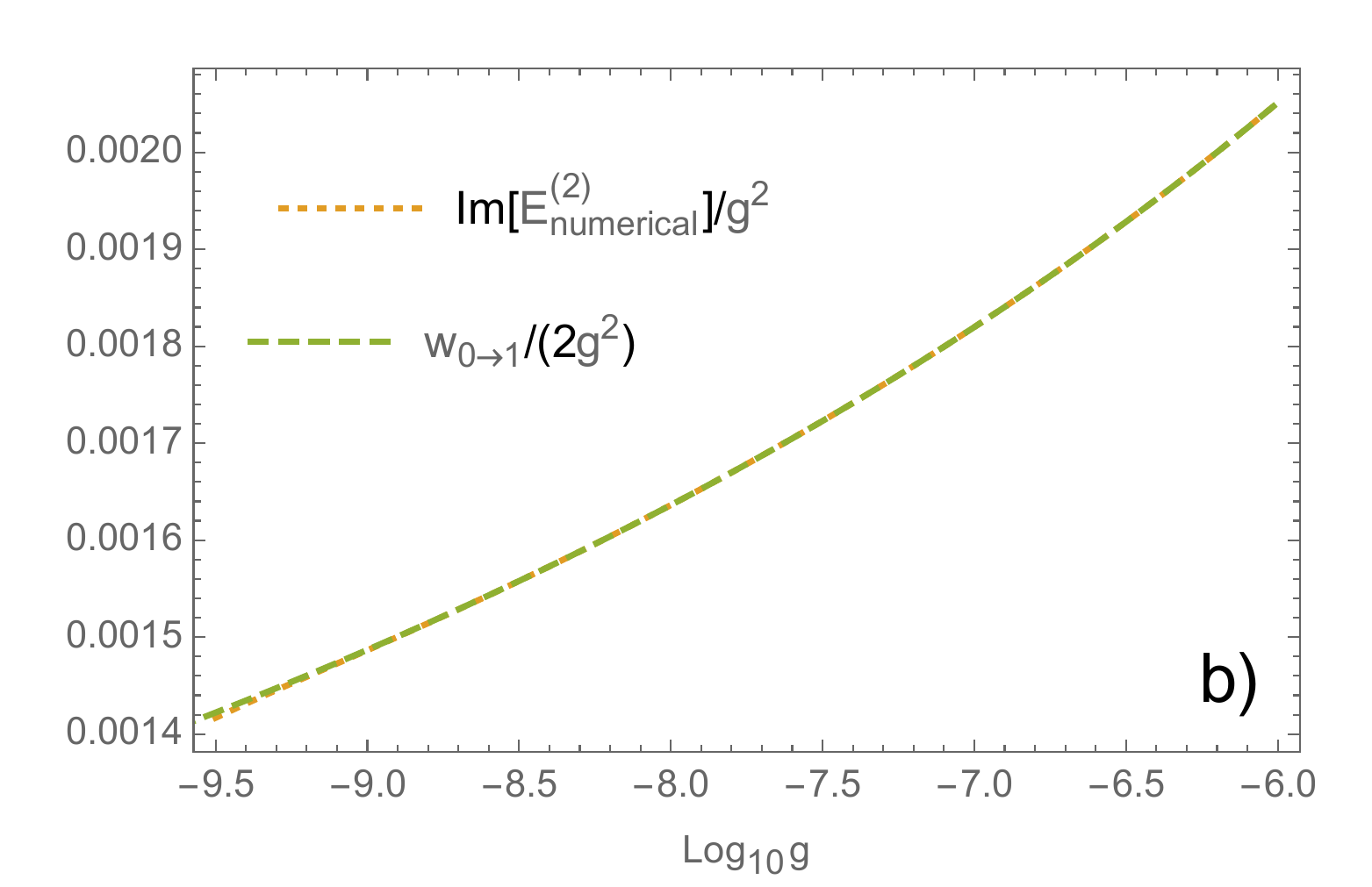}
	\caption{(Color online) (a)) The dependence on the coupling constant of the ratio of the exact numerical evaluation to the approximate analytical formula of the second iteration for the energy. The value $k_0$ in the analytical approximation is equal to $k_0 = \lambda\sqrt{3|\ln g|}$. (b)) The dependence of the imaginary part of the energy of the system on the coupling constant. The imaginary part corresponds to the finite lifetime of the state.}
	\label{fig1}
\end{figure}
In order to ensure that our interpretation is correct, we have evaluated the integrals numerically and have found in the limit of extremely small $g$ the ratio of the results via exact numerical versus analytical evaluations. This ratio is almost constant and is approximately equal to one, as presented in Figure~\ref{fig1}. Therefore, we can conclude that the main reason why conventional perturbation theory fails is related to the fact that the energy of the system is a non-analytical function of the coupling constant and consequently can not be expanded in a series over $g$ near a singular point.

The second interesting consequence of the numerical evaluation of the integral is related to the fact that the energy of the system contains a small imaginary part, which means that the state has a finite lifetime and is quasi-stationary. To prove this, we have calculated the transition probability to the state $|\Psi_{\vec P_1,1_{\vec k}}\ra$ for the case when a particle is at rest, i.e.
\begin{align}
	\frac{w_{0\rightarrow1}}{2} &= \pi \int |\la \Psi_{\vec P_1,   \{ n_{\vec k}\}}| \opa H |\Psi^{(L)}_{\vec P}\ra|^2 \delta\left(H_{\vec P_1, 1_{\vec k}; \vec P_1, 1_{\vec k}}-E_L^{(0)}\right)\frac{\Omega d\vec k}{(2\pi)^3} \nonumber
	\\
	&=\frac{\Omega}{2\pi}\frac{k^2}{|k+1+g^2 I^\prime_{\vec k}|}\frac{\left[u_{\vec k} \phi_{\vec k}^2\left(\frac{k^2}{2}+k\right)+\frac{g}{\sqrt{2 \Omega}}\frac{\phi_{\vec k} \phi_0}{\sqrt{k}}+u_{\vec k} \phi_{\vec k}^2 g^2 I_{\vec k} - E_L^{(0)} u_{\vec k} \phi_0^2\right]^2}{\phi_0^2 \phi_{\vec k}^2}\Bigg|_{\frac{k^2}{2}+k+g^2 I_{\vec k} - E_L^{(0)}=0}. \label{eq:second_order_iteration_for_the_energy_convergence_22}
\end{align}

\begin{figure}[t]
	\centering
		\includegraphics[scale=0.8]{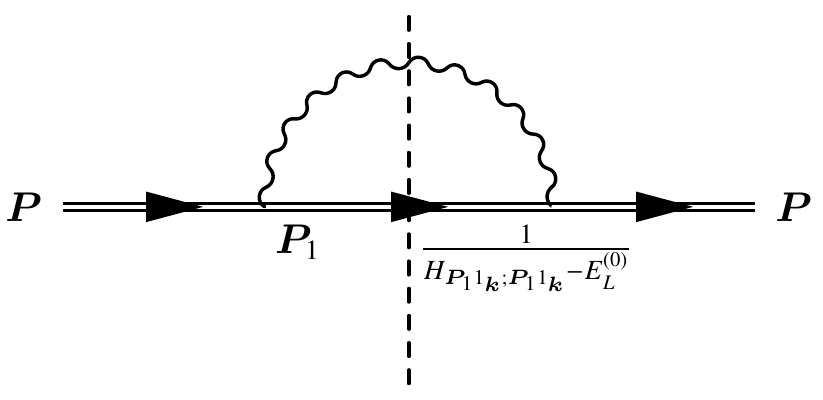}
	\caption{(Color online) Feynman diagram of the process. If the diagram is split by the dashed line, the imaginary part will correspond to the transition probability to the state $|\Psi_{\vec P_1,1_{\vec k}}\ra$. However, in the intermediate states the resolvent of the full Hamiltonian instead of a free one is used.}
	\label{fig2}
\end{figure}

The result of evaluation is presented in Figure~\ref{fig1}. As can be seen from the figure the two curves coincide exactly. This can be interpreted via the diagram technique \cite{LandauQED}. The second order iteration for the energy of the particle can be presented via the diagram depicted in Figure~\ref{fig2}. If the diagram is split by the dashed line, the imaginary part will correspond to the transition probability to the  state $|\Psi_{\vec P_1,1_{\vec k}}\ra$.

Here, we need to stress that contrary to standard perturbation theory in our formulation the conservation of energy is governed not by the free Hamiltonian $\opa H_0$, but through the expectation value of the total Hamiltonian $H_{\vec P_1 1_{\vec k};\vec P_1 1_{\vec k}}$. Therefore, for certain values of $k$ and for certain coupling constants $g$ this energy level might appear to be below the energy $E_L^{(0)}$, featuring a so called quasi-intersection of energy levels. If the transition probability to the state $|\Psi_{\vec P_1,1_{\vec k}}\ra$ were large, the description of the system with the state vectors (\ref{eq:iteration_scheme_basis_zero_order_approximation_to_the_energy_of_the_system_14}) would not be applicable and the reconstruction of the states would need to be performed, which takes into account the degeneracy between the energies $E_L^{(0)}$ and $H_{\vec P_1 1_{\vec k};\vec P_1 1_{\vec k}}$. In our case, however, the transition probability is small and consequently the description with a complex energy, with small imaginary part, is valid, in analogy to the theory of a natural line width of the atomic states or anharmonic oscillator $p^2/2+x^2/2-\mu x^4$, with $\mu>0$.

In order to conclude our formulation we have calculated the renormalized mass in the second iteration. In terms of the introduced abbreviations the second order iteration for the particle energy can be written as
\begin{align}
	E^{(2)}(\vec P,g) =  \frac{\frac{P^2}{2} + \tilde E_{L}^{(0)}(\vec P,g) + A_{\vec P}}{B_{\vec P}}, \label{eq:second_order_iteration_for_the_energy_convergence_24}
\end{align}
with
\begin{align}
	\tilde E_L^{(0)}(\vec P,g) &=  - \vec P\cdot\sum_{\vec m}\vec m |u_{\vec m}|^2 \frac{\phi_{\vec P - \vec m}^2}{\phi_{\vec P}^2} + \sum_{\vec m}\left(\frac{m^2}{2} + m\right)|u_{\vec m}|^2 \frac{\phi_{\vec P - \vec m}^2}{\phi_{\vec P}^2} + \frac{2g}{\sqrt{2 \Omega}}\sum_{\vec m}\frac{u_{\vec m}}{\sqrt{m}}\frac{\phi_{\vec P - \vec m}}{\phi_{\vec P}}.\label{eq:second_order_iteration_for_the_energy_convergence_25}
\end{align}

The quantity $A_{\vec P}$ reads as
\begin{align}
	A_{\vec P} &= \sum_{\vec k}\Bigg[\frac{P^2}{2}u_{\vec k}\left(\phi_{\vec P - \vec k}^2 - \phi_{\vec P}^2\right)+\left(\frac{k^2}{2}+k-\thp{P}{k}\right)u_{\vec k}\phi_{\vec P - \vec k}^2 + \frac{g}{\sqrt{2 \Omega}}\frac{\phi_{\vec P - \vec k}\phi_{\vec P}}{\sqrt{k}} + g^2 u_{\vec k} \phi_{\vec P - \vec k}^2 I_{\vec P - \vec k}-\tilde E_{L}^{(0)}(\vec P,g)u_{\vec k}\phi_{\vec P}^2\Bigg] \nonumber
	\\
	&\mspace{29mu}\times\left[\tilde E_{L}^{(0)}(\vec P,g)u_{\vec k}\phi_{\vec P-\vec k}^2-\left(\left(\frac{k^2}{2}+k-\thp{P}{k}\right)u_{\vec k}\phi_{\vec P - \vec k}^2 + \frac{g}{\sqrt{2 \Omega}}\frac{\phi_{\vec P - \vec k}\phi_{\vec P}}{\sqrt{k}} + g^2 u_{\vec k} \phi_{\vec P - \vec k}^2 I_{\vec P - \vec k}\right)\right]\frac{1}{\phi_{\vec P - \vec k}^2 \phi_{\vec P}^2} \nonumber
	\\
	&\mspace{29mu}\times \left[\left(\frac{k^2}{2}+k-\thp{P}{k}\right) + g^2 I_{\vec P-\vec k} - \tilde E_L^{(0)}(\vec P,g)\right]^{-1}, \label{eq:second_order_iteration_for_the_energy_convergence_26}
\end{align}
and the quantity $B_{\vec P}$ is equal to
\begin{align}
	B_{\vec P} &=1+ \sum_{\vec k}\left[\tilde E_{L}^{(0)}(\vec P,g)u_{\vec k}\phi_{\vec P-\vec k}^2-\left(\left(\frac{k^2}{2}+k-\thp{P}{k}\right)u_{\vec k}\phi_{\vec P - \vec k}^2 + \frac{g}{\sqrt{2 \Omega}}\frac{\phi_{\vec P - \vec k}\phi_{\vec P}}{\sqrt{k}} + g^2 u_{\vec k} \phi_{\vec P - \vec k}^2 I_{\vec P - \vec k}\right)\right]\frac{u_{\vec k}\left(\phi_{\vec P - \vec k}^2 - \phi_{\vec P}^2\right)}{\phi_{\vec P - \vec k}^2 \phi_{\vec P}^2} \nonumber
	\\
	&\mspace{29mu}\times \left[\left(\frac{k^2}{2}+k-\thp{P}{k}\right) + g^2 I_{\vec P-\vec k} - \tilde E_L^{(0)}(\vec P,g)\right]^{-1}. \label{eq:second_order_iteration_for_the_energy_convergence_27}
\end{align}

To proceed, we again break the limit of the integration into two parts, however, now we know that the main contribution to the energy of the system comes from the region $[0,k_0]$. In this region, we again drop all terms, with power of $g$ larger than one. We recall here that the classical component of the field $u_{\vec k}$ is proportional to $g$ and the energy $E_{L}^{(0)}(\vec P,g)\sim g^2$. In addition, the limit $\vec P\ll 1$ is considered. Moreover, as the quantity $B_{\vec P}$, after expansion over momentum $\vec P$, will have a form $B_{\vec P} = 1-g^2 F(P^2)$, we can set $\vec P=0$ in $B_{\vec P}$, in order to preserve the same accuracy.

In this approximation, the quantity $A_{\vec P}$ takes the form
\begin{align}
	A_{\vec P} = &-\frac{P^2}{2}\sum_{\vec k<\vec k_0}\left\{u_{\vec k}\frac{\left(\phi_{\vec P - \vec k}^2 - \phi_{\vec P}^2\right)}{\phi_{\vec P - \vec k}\phi_{\vec P}}\left[u_{\vec k}\frac{\phi_{\vec P - \vec k}}{\phi_{\vec P}} + \frac{g}{\sqrt{2 \Omega}}\frac{1}{\sqrt{k}} \left(\frac{k^2}{2}+k-\thp{P}{k}\right) ^{-1}\right]\right\} \nonumber
	\\
	&-\sum_{\vec k<\vec k_0}\left[\left(\frac{k^2}{2}+k-\thp{P}{k}\right)u_{\vec k}\frac{\phi_{\vec P - \vec k}}{\phi_{\vec P}} + \frac{g}{\sqrt{2 \Omega}}\frac{1}{\sqrt{k}} \right]^2\left(\frac{k^2}{2}+k-\thp{P}{k}\right)^{-1} \label{eq:second_order_iteration_for_the_energy_convergence_28}
\end{align}
and
\begin{align}
	B_{\vec P} = B_0 = 1 -\sum_{\vec k < \vec k_0}\frac{\left(u_{\vec k} \frac{\phi_{\vec k}^2}{\phi_0^2}\left(\frac{k^2}{2}+k\right)+\frac{g}{\sqrt{2 \Omega}}\frac{\phi_{\vec k}}{\phi_0\sqrt{k}}\right)u_{\vec k}(\phi_{\vec k}^2-\phi_0^2)}{\phi_{\vec k}^2(\frac{k^2}{2}+k)}. \label{eq:second_order_iteration_for_the_energy_convergence_29}
\end{align}

If the definition of $E_{L}^{(0)}(\vec P,g)$ (\ref{eq:second_order_iteration_for_the_energy_convergence_25}), together with equations (\ref{eq:second_order_iteration_for_the_energy_convergence_28}) and (\ref{eq:second_order_iteration_for_the_energy_convergence_29}), is used in equation (\ref{eq:second_order_iteration_for_the_energy_convergence_24}), the second iteration for the energy of a moving particle is obtained
\begin{align}
	E^{(2)}(\vec P,g) &= E^{(2)}(0,g) -\frac{g^2}{2 \Omega}\sum_{\vec k<\vec k_0}\frac{(\thp{P}{k})^2}{k(k^2/2+k)^3} \nonumber
	\\
	&\mspace{50mu}+\frac{P^2}{2}\left\{1-\sum_{\vec k < \vec k_0} \left[u_{\vec k}^2 \left(\frac{\phi_{\vec k}^2}{\phi_{0}^2}-1\right)+\frac{g}{\sqrt{2 \Omega}}\frac{u_{\vec k}}{\sqrt{k}(k^2/2+k)}\frac{\left(\phi_{\vec k}^2 - \phi_{0}^2\right)}{\phi_{\vec k}\phi_{0}}\right]\right\}\nonumber
	\\
	&\mspace{50mu}\times\left(1-\sum_{\vec k < \vec k_0} \left[u_{\vec k}^2 \left(\frac{\phi_{\vec k}^2}{\phi_{0}^2}-1\right)+\frac{g}{\sqrt{2 \Omega}}\frac{u_{\vec k}}{\sqrt{k}(k^2/2+k)}\frac{\left(\phi_{\vec k}^2 - \phi_{0}^2\right)}{\phi_{\vec k}\phi_{0}}\right]\right)^{-1},\label{eq:second_order_iteration_for_the_energy_convergence_30}
\end{align}
or after simplification, taking into account the fact that the sum in the denominator is proportional to $g^2$, one finally obtains
\begin{align}
	E^{(2)}(\vec P,g) &= E^{(2)}(0,g) +\frac{P^2}{2}-\frac{g^2}{2 \Omega}\sum_{\vec k<\vec k_0}\frac{(\thp{P}{k})^2}{k(k^2/2+k)^3}.\label{eq:second_order_iteration_for_the_energy_convergence_31}
\end{align}

From here, we see that the second iteration for the renormalized mass 
\begin{align}
	m^{(2)*} \approx 1+\frac{g^2}{6\pi^2}\label{eq:second_order_iteration_for_the_energy_convergence_32}
\end{align}
coincides with the one via perturbation theory.

\section{Conclusion} 
\label{sec:conclusion}
In current methods of renormalization in QFT, the momentum cut-off plays an important role \cite{Collins1984}, which in fact is an additional and undefined parameter of the theory. Usually, the inclusion of such parameter for a concrete model is justified with the argument that the theory becomes incorrect on a small scale, where a more general theory must be used instead. For example, in the case of QED it is widely accepted that on a small scale the Standard Model, with its own characteristic length, should be rather used. However, in the Standard Model, as in its possible generalizations, for the renormalization of perturbation theory the cut-off is required. Consequently, we come to the requirement of the inclusion of some ``fundamental length'' or unobservable parameter of any QFT. 

However, the Fr\"ohlich Hamiltonian demonstrates the absence of a cut-off in the polaron theory. In this QFT all corrections are determined through convergent integrals and, consequently, the cut-off is not required. Here we considered a more general QFT than the one associated with the polaron problem, for which standard perturbation theory gives rise to divergences. The main result of the present work consists in the construction of a calculation scheme for this more general QFT that only leads to convergent integrals. In addition to that, the regularization of all integrals is related to the effective-cut-off momentum, which is defined through the parameters of the system itself. Moreover, the divergences of standard perturbation theory are explained through the energy being a non-analytical function of the coupling constant, of a form $\ln(\sqrt{|\ln g|}/2+1)$, around zero, and, therefore, can not be represented as a power series around this singular point. It is also important that the character of the singularity, defined in equation (\ref{eq:second_order_iteration_for_the_energy_convergence_20}) in the weak coupling limit does not depend on the particular choice of the wave functions $\phi_{\vec P}(\vec r)$ of the zeroth-oder approximation.

From a formal point of view, the convergence of all integrals is explained as follows: i) the use of the decomposition (\ref{eq:iteration_scheme_basis_zero_order_approximation_to_the_energy_of_the_system_2}), i.e. the special state vectors, which are the product of  the wave function of a localized particle and a coherent state of the field and ii) the calculation of the energy of the system with the iteration scheme (\ref{eq:iteration_scheme_basis_zero_order_approximation_to_the_energy_of_the_system_10}), in which the resolvent of the operator $[H_{kk}-E^{(0)}]^{-1}$ contains the matrix elements of the full Hamiltonian of the system. In standard perturbation theory the Hamiltonian of non-interacting fields is used in the analogous expressions.

From a physical point of view the argument i) corresponds to avoiding an adiabatic switch off of the interaction. This means that a particle during its existence time is considered as ``dressed'', i.e. to be in a localized state which is created due to the interaction between the particle and the field. The argument ii) leads to the ``cutting'' of all integrals for a large momentum due to the reconstruction of a localized state in intermediate states, caused by the quasi-intersection of the ground and the single-phonon states.

Our approach should not be considered and does not pretend to be the full solution of the renormalization problem in QFT, specifically, because of the use of a simple, non-relativistically covariant model. Nevertheless, it demonstrates an alternative, succeeding without introducing any phenomenological momentum cut-off.

\begin{acknowledgments}
The authors are grateful to S. Cavaletto for useful discussions.	
\end{acknowledgments}

\section*{Appendix A: Proof that the states (\ref{eq:iteration_scheme_basis_zero_order_approximation_to_the_energy_of_the_system_14}) are eigenstates of the total momentum operator} 
\label{sec:appendix}
In this appendix we present an explicit proof that the total momentum operator 
\begin{align*}
	\opA P = -\ri \nabla_{\vec r} + \sum_{\vec k}\vec k \opa a_{\vec k}^\dag \opa a_{\vec k}
\end{align*}
commutes with the Hamiltonian 
\begin{align*}
	\opa H = -\frac{1}{2}\Delta + \sum_{\vec k}k \opa a_{\vec k}^\dag \opa a_{\vec k} + \frac{g}{\sqrt{2\Omega}}\sum_{\vec k}A_{\vec k} \left(e^{\ri\thp{k}{r}}\opa a_{\vec k}+e^{-\ri\thp{k}{r}}\opa a^\dag_{\vec k}\right)
\end{align*}
of the system and that the states (\ref{eq:iteration_scheme_basis_zero_order_approximation_to_the_energy_of_the_system_14}) are eigenstates of $\opA P$
\begin{align}
	\opA P |\Psi^{(0)}_{\vec P_1,n_{\vec k}}\ra &= \vec P_1 |\Psi^{(0)}_{\vec P_1,n_{\vec k}}\ra,\label{A1}
\end{align}
consequently forming a complete set in the Hilbert space.

Let us begin with the commutator:
\begin{align}
	[\opa H, \opA P] &= \left[\frac{g}{\sqrt{2\Omega}}\sum_{\vec k}A_{\vec k} \left(e^{\ri\thp{k}{r}}\opa a_{\vec k}+e^{-\ri\thp{k}{r}}\opa a^\dag_{\vec k}\right), -\ri \nabla\right] \nonumber
	\\
	&\mspace{90mu}+\left[\frac{g}{\sqrt{2\Omega}}\sum_{\vec k}A_{\vec k} \left(e^{\ri\thp{k}{r}}\opa a_{\vec k}+e^{-\ri\thp{k}{r}}\opa a^\dag_{\vec k}\right),\sum_{\vec k}\vec k \opa a_{\vec k}^\dag \opa a_{\vec k}\right] \nonumber
	\\
	&=\frac{g}{\sqrt{2 \Omega}}\sum_{\vec k}A_{\vec k}\left(\opa a_{\vec k}e^{\ri \thp{k}{r}}(-\vec k)+\opa a_{\vec k}^\dag e^{-\ri \thp{k}{r}}\vec k\right) \nonumber
	\\
	&\mspace{90mu}+\frac{g}{\sqrt{2 \Omega}}\sum_{\vec k}A_{\vec k}\left(\opa a_{\vec k}e^{\ri \thp{k}{r}}\vec k+\opa a_{\vec k}^\dag e^{-\ri \thp{k}{r}}(-\vec k)\right) = 0,\label{A2}
\end{align}
which was to be proven.

Now we will demonstrate that the relation (\ref{A1}) holds. Also, we will introduce the notations
\begin{align}
	\opa D(\vec R) &= \exp\left\{\sum_{\vec q}(u_{\vec q} e^{- \ri\vec q \cdot \vec R}\opa a_{\vec q}^\dag - u_{\vec q}^* e^{ \ri\vec q \cdot \vec R}\opa a_{\vec q})\right\}, \text{ with} \label{A3}
	\\
	&\opa D^\dag(\vec R)\opa D(\vec R) = \opa D(\vec R)\opa D^\dag(\vec R) = 1,\label{A4}
	\\
	&\opa D^\dag(\vec R)\opa a_{\vec k}\opa D(\vec R) = \opa a_{\vec k} +u_{\vec k}e^{-\ri\thp{k}{R}},\label{A5}
	\\
	&\opa D^\dag(\vec R)\opa a^\dag_{\vec k}\opa D(\vec R) = \opa a^\dag_{\vec k} + u^*_{\vec k}e^{\ri\thp{k}{R}},\label{A6}
	\\
	&\ri\frac{\partial \opa D(\vec R)}{\partial \vec R} = \opa D(\vec R)\sum_{\vec q}\vec q\left(u_{\vec q}e^{-\ri\thp{q}{R}}\opa a_{\vec q}^\dag+u^*_{\vec q}e^{\ri\thp{q}{R}}\opa a_{\vec q}\right)\label{A7}
\end{align}

Consequently, with the help of equations (\ref{A3}-\ref{A7}) we may write
\begin{align}
	\opA P |\Psi^{(0)}_{\vec P_1,n_{\vec k}}\ra &= \left(-\ri \nabla_{\vec r} + \sum_{\vec q}\vec q \opa a_{\vec q}^\dag \opa a_{\vec q}\right)\frac{1}{N_{\vec P_1,n_{\vec k}}\sqrt{\Omega}}\int d \vec R \phi_{\vec P_1}(\vec r - \vec R)\exp\left\{\ri(\vec P_1 - \vec k n_{\vec k})\cdot \vec R \right\}\opa D(\vec R)|n_{\vec k}\ra \nonumber
	\\
	&=\frac{1}{N_{\vec P_1,n_{\vec k}}\sqrt{\Omega}}\int d \vec R (-\ri \nabla_{\vec r})(\phi_{\vec P_1}(\vec r - \vec R))\exp\left\{\ri(\vec P_1 - \vec k n_{\vec k})\cdot \vec R \right\}\opa D(\vec R)|n_{\vec k}\ra \nonumber
	\\
	&+\frac{1}{N_{\vec P_1,n_{\vec k}}\sqrt{\Omega}}\int d \vec R \phi_{\vec P_1}(\vec r - \vec R)\exp\left\{\ri(\vec P_1 - \vec k n_{\vec k})\cdot \vec R \right\}\sum_{\vec q}\vec q \opa a_{\vec q}^\dag \opa a_{\vec q}\opa D(\vec R)|n_{\vec k}\ra. \label{A8}
\end{align}
By noticing that $-\ri \nabla_{\vec r}(\phi_{\vec P_1}(\vec r - \vec R)) = \ri \nabla_{\vec R}(\phi_{\vec P_1}(\vec r - \vec R))$ and transforming $\sum_{\vec q}\vec q \opa a_{\vec q}^\dag \opa a_{\vec q}\opa D(\vec R) = \opa D(\vec R)\opa D^\dag(\vec R)\sum_{\vec q}\vec q \opa a_{\vec q}^\dag \opa a_{\vec q}\opa D(\vec R)$ one obtains
\begin{align}
	\opA P |\Psi^{(0)}_{\vec P_1,n_{\vec k}}\ra &= \frac{1}{N_{\vec P_1,n_{\vec k}}\sqrt{\Omega}}\int d \vec R (\ri \nabla_{\vec R})\left[\phi_{\vec P_1}(\vec r - \vec R)\exp\left\{\ri(\vec P_1 - \vec k n_{\vec k})\cdot \vec R \right\}\opa D(\vec R)\right]|n_{\vec k}\ra \nonumber
	\\
	&-\frac{1}{N_{\vec P_1,n_{\vec k}}\sqrt{\Omega}}\int d \vec R \phi_{\vec P_1}(\vec r - \vec R)(\ri \nabla_{\vec R})\left[\exp\left\{\ri(\vec P_1 - \vec k n_{\vec k})\cdot \vec R \right\}\opa D(\vec R)\right]|n_{\vec k}\ra \nonumber
	\\
	&+\frac{1}{N_{\vec P_1,n_{\vec k}}\sqrt{\Omega}}\int d \vec R \phi_{\vec P_1}(\vec r - \vec R)\exp\left\{\ri(\vec P_1 - \vec k n_{\vec k})\cdot \vec R \right\}\opa D(\vec R) \nonumber
	\\
	&\mspace{350mu}\times\sum_{\vec q}\vec q \left(\opa a_{\vec q}^\dag+u_{\vec q}^* e^{\ri\thp{q}{R}}\right) \left(\opa a_{\vec q}+u_{\vec q} e^{-\ri\thp{q}{R}}\right)|n_{\vec k}\ra. \label{A9}
\end{align}
The first term in equation (\ref{A9}) vanishes due to the square-integrability of the function $\phi_{\vec P_1}(\vec r-\vec R)$. The derivative in the second term is equal to
\begin{align}
	(\ri \nabla_{\vec R})\left[\exp\left\{\ri(\vec P_1 - \vec k n_{\vec k})\cdot \vec R \right\}\opa D(\vec R)\right] &= -(\vec P_1 - \vec k n_k)\exp\left\{\ri(\vec P_1 - \vec k n_{\vec k})\cdot \vec R \right\}\opa D(\vec R) \nonumber
	\\
	&+\exp\left\{\ri(\vec P_1 - \vec k n_{\vec k})\cdot \vec R \right\}\opa D(\vec R)\sum_{\vec q}\vec q\left(u_{\vec q}e^{-\ri\thp{q}{R}}\opa a_{\vec q}^\dag+u^*_{\vec q}e^{\ri\thp{q}{R}}\opa a_{\vec q}\right) \label{A10}
\end{align}
and, therefore, the terms which are not proportional to $\vec P_1$ in equation (\ref{A10}) cancel the last term in equation (\ref{A9}). As a result, equation (\ref{A9}) transforms into
\begin{align}
	\opA P |\Psi^{(0)}_{\vec P_1,n_{\vec k}}\ra = \vec P_1 |\Psi^{(0)}_{\vec P_1,n_{\vec k}}\ra, \label{A11}
\end{align}
which was to be proven. 

According to reference \cite{landau1965quantum}, the eigenstates of a Hermitian operator form a complete and orthogonal set of functions in the Hilbert space. As the functions (\ref{eq:iteration_scheme_basis_zero_order_approximation_to_the_energy_of_the_system_14}) are eigenstates of the Hermitian operator $\opA P$, they form a complete orthogonal set for arbitrary generalized parameters $\phi_{\vec P_1}(\vec r - \vec R)$ and $u_{\vec k}$. 

\section*{Appendix B: Matrix elements calculation} 
\label{sec:matrix_elements_calculation}
In all subsequent calculations, the matrix elements of a type
\begin{align}
	\label{1}
	\la n_{\vec j}|&\exp\left(-\sum_{\vec m}\opa a_{\vec m}^\dag u_{\vec m}e^{-\ri\thp{m}{R^\prime}}-\opa a_{\vec m}u^*_{\vec m}e^{\ri\thp{m}{R^\prime}}\right)\sum_{\vec l}f(\opa a_{\vec l},\opa a^\dag_{\vec l})\nonumber
	\\
	&\times\exp\left(\sum_{\vec m}\opa a_{\vec m}^\dag u_{\vec m}e^{-\ri\thp{m}{R}}-\opa a_{\vec m} u^*_{\vec m}e^{\ri\thp{m}{R}}\right)|n_{\vec k}\ra
\end{align}
need to be evaluated. By using the identities
\begin{align}
	\opa D &= e^{\beta \opa a^\dag - \beta^* \opa a} = e^{-|\beta|^2/2}e^{\beta \opa a^\dag}e^{-\beta^* \opa a} = e^{|\beta|^2 /2}e^{-\beta^* \opa a}e^{\beta \opa a^\dag}, \label{2}
	\\
	\opa D^{-1}\opa a \opa D &= \opa a +\beta, \quad \opa D^{-1}\opa a^\dag \opa D= \opa a^\dag + \beta^*,\label{3}
\end{align}
equation (\ref{1}) can be transformed into the form
\begin{align}
	&\frac{\exp(\sum_{\vec m}|u_{\vec m}|^2 (e^{-\ri\vec m \cdot (\vec R - \vec R^\prime)}-1))}{\sqrt{n_{\vec k}!n_{\vec j}!}}\nonumber
	\\
	&\times\la0|(\opa a_{\vec j} - u_{\vec j}e^{-\ri\thp{j}{R^\prime}}+u_{\vec j}e^{-\ri\thp{j}{R}})^{n_{\vec j}}\sum_{\vec l}f\left(\opa a_{\vec l}+u_{\vec l} e^{-\ri\thp{l}{R}},\opa a^\dag_{\vec l}+u_{\vec l}^* e^{\ri\thp{l}{R^\prime}}\right)\nonumber
	\\
	&\times(\opa a_{\vec k}^\dag + u_{\vec k}^* e^{\ri\thp{k}{R^\prime}}-u_{\vec k}^* e^{\ri\thp{k}{R}})^{n_{\vec k}}|0\ra. \label{4}
\end{align}

The evaluation of equation (\ref{4}) is performed in the usual manner, i.e, by noticing that $\opa a|0\ra = \la0|\opa a^\dag = 0$ and the vacuum average is not equal to zero only if the number of creation operators is equal to the one of annihilation operators and is an even number.
\section*{Appendix C: Ground state energy} 
\label{sec:ground_state_energy}
According to equation (\ref{eq:iteration_scheme_basis_zero_order_approximation_to_the_energy_of_the_system_18}) of the manuscript, the ground state energy	is defined as
\begin{align}
	E_L^{(0)} = \la \Psi^{(L)}_{\vec P}|\opa H|\Psi^{(L)}_{\vec P}\ra \label{5}
\end{align}
with the wave function
\begin{align}
	\label{6}
	|\Psi^{(L)}_{\vec P}\ra   &= \frac{1}{N_{\vec P}\sqrt{\Omega}} \int d{\vec R}\,\phi_{\vec P}({\vec r}-{\vec R})\exp \left(\ri\thp{P}{R}+   \sum_{\vec k}\left(u_{\vec k}\opa a^\dag_{\vec k}e^{-\ri\thp{k}{R}}-\frac{1}{2}u^2_{\vec k} \right)\right)| 0 \ra
\end{align}
and Hamiltonian 
\begin{align}
	\opa H &= \frac{1}{2}\left(\opA{P}^2 -2 \sum_k \opa a_k^\dag \opa a_k \vec k \cdot \opA{P}+\left(\sum_k \opa a_k^\dag \opa a_k \vec k\right)^2\right)+\sum_{\vec k}\omega_k \opa a^+_{\vec k}\opa a_{\vec k}\nonumber
	\\
	&+ \frac{g}{\sqrt{\Omega}}\sum_{\vec k}\frac{1}{\sqrt{2\omega_k}} \left(e^{\ri\vec k\vec r}\opa a_{\vec k}+e^{-\ri\vec k\vec r}\opa a^+_{\vec k}\right). \label{7}
\end{align}
The normalization constant $N_{\vec P}$ is found from the condition
\begin{align}
	\label{8}
	\la\Psi^{(L)}_{\vec P}|\Psi^{(L)}_{\vec P}\ra = 1.
\end{align}
In order to evaluate equation (\ref{8}), we use equation (\ref{4}), in which $n_{\vec k} = n_{\vec j} = 0$ and $f(\opa a_{\vec l},\opa a_{\vec l}^\dag) = \delta_{\vec l,0}$. This gives immediately the result 
\begin{align}
	|N_{\vec P}|^2 = \int d{\vec  R}\int d{\vec r}\,\phi_{\vec P}^*({\vec r})\phi_{\vec P}({\vec r}-{\vec R}) \exp\left(\sum_{\vec
		k}|u_{\vec k}|^2(e^{-\ri\thp{k}{R}}-1) + \ri \thp{P}{R}\right).\label{9}
\end{align}

The expectation value of the energy is performed in exactly the same way. First of all, the matrix elements of the field states are calculated with the help of equation (\ref{4}). For example, if the function $f$ is chosen as $f = \vec l \opa a_{\vec l}^\dag \opa a_{\vec l}$, $n_{\vec k} = n_{\vec j} = 0$, we immediately find
\begin{align}
	\vec Q &= \la \sum_{\vec l}\vec l \opa a_{\vec l}^\dag \opa a_{\vec l}\ra \nonumber
	\\
	&=\int d\vec R d\vec R^\prime d\vec r \phi^*(\vec r- \vec R^\prime)\phi(\vec r - \vec R) \sum_{\vec l}\vec l |u_{\vec l}|^2 \nonumber
	\\
	&\times\exp\left(\sum_{\vec m}|u_{\vec m}|^2 (e^{-\ri\vec m \cdot (\vec R - \vec R^\prime)}-1)+\ri(\vec P - \vec l)\cdot(\vec R - \vec R^\prime)\right).\label{10}
\end{align} 

Then by carrying out the change of variables $\vec R - \vec R^\prime = \vec R_1$ and $\vec r - \vec R^\prime = \vec \rho$, we obtain 
\begin{align}
	\vec Q &= \frac{1}{|N_{\vec P}|^2} \sum_{\vec k}\vec k |u_{\vec k}|^2\int d{\vec  R} d{\vec r}\,\phi_{\vec P}^*({\vec r})\phi_{\vec P}({\vec r - \vec R}) e^{\Phi (\vec R) + \ri (\vec P - \vec k)\cdot \vec R},\label{11}
	\\
	\Phi (\vec R)&= \sum_{\vec k}|u_{\vec k}|^2(e^{-\ri\thp{k}{R}}-1). \nonumber
\end{align}

All other matrix elements are evaluated in exactly the same fashion. Consequently, we obtain expression (\ref{eq:iteration_scheme_basis_zero_order_approximation_to_the_energy_of_the_system_20}) of the manuscript:
\begin{align}
	&E_{L}^{(0)}(\vec P,g)= \frac{P^2}{2} - \thp{P}{Q} + G + E_{\text{f}}(\vec P) + E_{\text{int}}(\vec P); \label{12}
	\\
	&\vec Q = \frac{1}{|N_{\vec P}|^2} \sum_{\vec k}\vec k |u_{\vec k}|^2\int d{\vec  R} d{\vec r}\,\phi_{\vec P}^*({\vec r})\phi_{\vec P}({\vec r - \vec R}) e^{\Phi (\vec R) + \ri (\vec P - \vec k)\cdot \vec R}; \nonumber
	\\
	& G =\frac{1}{2} \frac{1}{|N_{\vec P}|^2}\sum_{\vec m,\vec l}\thp{m}{l}|u_{\vec m}|^2 |u_{\vec l}|^2\int d\vec r d\vec R \phi^*(\vec r)\phi(\vec r - \vec R)e^{\ri\thp{P}{R}+\Phi(\vec R)-\ri(\vec m + \vec l)\cdot R};\nonumber
	\\
	&E_{\text{f}}(\vec P) = \frac{1}{|N_{\vec P}|^2} \sum_{\vec k}\left(k + \frac{k^2}{2}\right)|u_{\vec k}|^2\int d{\vec  R} d{\vec r}\,\phi_{\vec P}^*({\vec r})\phi_{\vec P}({\vec r-\vec R}) e^{\Phi (\vec R) + \ri (\vec P - \vec k)\cdot \vec R};\nonumber
	\\
	&E_{\text{int}}(\vec P) = \frac{g}{|N_{\vec P}|^2 } \sum_{\vec k}\frac{u_{\vec k}}{\sqrt{2 k \Omega}} \int d{\vec  R} d{\vec r}\left(\phi_{\vec P}^*({\vec r}+\vec R)\phi_{\vec P}({\vec r})+\phi_{\vec P}^*({\vec r})\phi_{\vec P}({\vec r}-{\vec R})\right)e^{\Phi (\vec R) + \ri (\thp{P}{R} + \thp{k}{r})};\nonumber
	\\
	&\Phi (\vec R)= \sum_{\vec k}|u_{\vec k}|^2(e^{-\ri\thp{k}{R}}-1);\nonumber
	\\
	&|N_{\vec P}|^2 = \int d{\vec  R} d{\vec r}\,\phi_{\vec P}^*({\vec r})\phi_{\vec P}({\vec r}-{\vec R}) e^{\Phi(\vec R) + \ri \thp{P}{R}}.\nonumber
\end{align}

We notice here one more time that the Fourier component of the function reads
\begin{align}
	\phi(\vec r) = \frac{\lambda^{\frac{3}{2}}}{\pi^{\frac{3}{4}}}e^{-\frac{\lambda^2 r^2}{2}}. \label{13}
\end{align}
and the classical component of the field look like
\begin{align}
	u_{\vec k} &= -\frac{g}{\sqrt{2 \Omega}}\frac{1}{\sqrt{k^3}}\int d\vec r |\phi(\vec r)|^2 e^{-\ri\thp{k}{r}} = -\frac{g}{\sqrt{2 \Omega}}\frac{e^{-\frac{k^2}{4 \lambda^2}}}{\sqrt{k^3}}; \label{14}
	\\
	\phi_{\vec k} &= \int d\vec r \phi(\vec r)e^{-\ri\thp{k}{r}} = 2\sqrt{2}\frac{\pi^{\frac{3}{4}}}{\lambda^{\frac{3}{2}}} e^{-\frac{k^2}{2 \lambda^2}} = \phi_0 e^{-\frac{k^2}{2 \lambda^2}}. \label{15}
\end{align}
In order to calculate the energy, we firstly neglect the function
\begin{align}
	\Phi(R) = \sum_{\vec k}|u_{\vec k}|^2 (e^{-\ri\thp{k}{R}}-1)=g^2 \frac{1}{4\pi^2}\int_0^\infty dt \frac{e^{-\frac{t^2}{2 }}}{t}\left(\frac{\sin \lambda R t}{\lambda R t}-1\right)\sim g^2 \label{16}
\end{align}
in equation (\ref{12}). The remaining quantities can be rewritten employing the Fourier transform of the function $\phi(\vec r)$. As for example
\begin{align}
	\int d\vec R_1 d \vec \rho \phi^*(\vec \rho)\phi(\vec \rho - \vec R_1) e^{-\ri\thp{k}{R_1}} =\int d \vec \rho \phi^*(\vec \rho)e^{-\ri\thp{k}{\rho}}\int d \vec R \phi(\vec R)e^{\ri\thp{k}{R}} = \phi^*_{\vec k}\phi_{-\vec k} &= \phi_{\vec k}^2, \label{17}
	\\
	\int d\vec R_1 d \vec \rho \phi^*(\vec \rho)\phi(\vec \rho - \vec R_1) = |\phi_0|^2 &= \phi_0^2, \label{18}
	\\
	\int d \vec R_1 d \vec \rho \phi^*(\vec \rho)\phi(\vec \rho - \vec R_1) e^{-\ri\thp{k}{\rho}} = \phi^*_{\vec k}\phi_0 &= \phi_{\vec k}\phi_0, \label{19}
\end{align}
and by plugging equations (\ref{17}-\ref{19}) into equation (\ref{12}), one finds
\begin{align}
	E^{(0)}_L(0, g) = \frac{1}{2}\sum_{\vec m,\vec l}\thp{m}{l}|u_{\vec m}|^2 |u_{\vec l}|^2\,\frac{\phi_{\vec l+ \vec m}^2}{\phi_0^2}+\sum_{\vec k}\left(k + \frac{k^2}{2}\right)|u_{\vec k}|^2 \frac{\phi_{\vec k}^2}{\phi_0^2} + \frac{2g}{\sqrt{2 \Omega}}\sum_{\vec k}\frac{u_{\vec k}}{\sqrt{k}}\frac{\phi_{\vec k}}{\phi_0}. \label{20}
\end{align}

By insertion of the definitions of the classical component of the field $u_{\vec k}$ and the Fourier transform of the function $\phi_{\vec k}$, defined in equations (\ref{14}) and (\ref{15}), we find
\begin{align}
	E^{(0)}_L&(0, g) = \frac{g^4}{8(2\pi)^6}\int d\vec l d\vec m \frac{\thp{m}{l}}{m^3 l^3}e^{-\frac{3}{2}\frac{m^2}{\lambda^2}-\frac{3}{2}\frac{l^2}{\lambda^2}-\frac{2\thp{m}{l}}{\lambda^2}}+\frac{g^2}{2(2\pi)^3}\int \frac{d\vec k}{k^2}\left(1+\frac{k}{2}\right)e^{-\frac{3}{2}\frac{k^2}{\lambda^2}} \nonumber
	\\
	&- \frac{g^2}{(2\pi^3)}\int \frac{d\vec k}{k^2}e^{-\frac{3}{4}\frac{k^2}{\lambda^2}} =\frac{g^4}{8(2\pi)^5}\int \frac{d\vec l}{l^2}e^{-\frac{3}{2}\frac{l^2}{\lambda^2}}\int dm e^{-\frac{3}{2}\frac{m^2}{\lambda^2}}\frac{-2lm \lambda^2 \cosh\frac{2lm}{\lambda^2}+\lambda^4\sinh\frac{2lm}{\lambda^2}}{2l^2m^2} \nonumber
	\\
	&+\frac{g^2}{24\pi^2}\left(\lambda(-4+\sqrt{2})\sqrt{3\pi}+\lambda^2\right) =\frac{g^4 \lambda^2}{16(2\pi)^4}\int_0^\infty \frac{du}{u^2}\left(4u-e^{\frac{2}{3}u^2}\sqrt{6\pi}\text{Erf}\left(\sqrt{\frac{2}{3}}u\right)\right)e^{-\frac{3}{2}u^2}\nonumber
	\\
	&+\frac{g^2}{24\pi^2}\left(\lambda(-4+\sqrt{2})\sqrt{3\pi}+\lambda^2\right) \nonumber
	\\
	&=-\frac{g^4 \lambda^2}{2^8 \pi^4}\alpha + \frac{g^2}{24\pi^2}\left(\lambda(-4+\sqrt{2})\sqrt{3\pi}+\lambda^2\right),\label{21}
\end{align}
where $\alpha = 0.736559$. 

To find $\lambda$, we minimize the energy, which results in the equation
\begin{align}
	\frac{\partial E^{(0)}_L(0, g)}{\partial \lambda} = -\frac{2g^4 \lambda }{2^8 \pi^4}\alpha + \frac{g^2}{24\pi^2}\left((-4+\sqrt{2})\sqrt{3\pi}+2\lambda\right),\label{22}
\end{align}
from here we find
\begin{align}
	\lambda = -\frac{\sqrt{3\pi}}{2}(-4+\sqrt{2})\frac{1}{1-\frac{3 \alpha g^2}{32\pi^2}}\approx -\frac{\sqrt{3\pi}}{2}(-4+\sqrt{2})\left(1+\frac{3 \alpha g^2}{32\pi^2}\right)\label{23}
\end{align}
and by plugging $\lambda$ in equation (\ref{21})
\begin{align}
	E^{(0)}_L(0, g) = -g^2\frac{(-4+\sqrt{2})^2}{32\pi} - \frac{3 \alpha g^4(-4+\sqrt{2})^2}{2^{10}\pi^3} + O(g^6). \label{24}
\end{align}
\section*{Appendix D: Mass renormalization in zeroth-order approximation} 
\label{sec:mass_renormalization_in_zeroth_order_approximation}
In the weak coupling limit we find the renormalized mass in zeroth-order approximation. For this purpose, we rewrite the energy through Fourier components for the case $\vec P\neq0$. This yields
\begin{align}
	E^{(0)}_L(\vec P, g)&= \frac{P^2}{2} - \thp{P}{Q} + G + E_{\text{f}}(\vec P) + E_{\text{int}}(\vec P); \label{25}
	\\
	\vec Q &= \frac{1}{|N_{\vec P}|^2} \sum_{\vec k}\vec k |u_{\vec k}|^2 \phi_{\vec P - \vec k}^2; \label{26}
	\\
	G &= \frac{1}{2} \frac{1}{|N_{\vec P}|^2} \sum_{\vec m,\vec l}\thp{m}{l}|u_{\vec m}|^2 |u_{\vec l}|^2 \phi_{\vec P - \vec l - \vec m}^2;\label{27}
	\\
	E_{\text{f}}(\vec P) &= \frac{1}{|N_{\vec P}|^2} \sum_{\vec k}\left(k + \frac{k^2}{2}\right)|u_{\vec k}|^2 \phi_{\vec P - \vec k}^2; \label{28}
	\\
	E_{\text{int}}(\vec P) &= \frac{g}{|N_{\vec P}|^2 } \sum_{\vec k}\frac{u_{\vec k}}{\sqrt{2 k \Omega}} (\phi_{\vec P}\phi_{\vec P -\vec k}+\phi_{\vec P} \phi_{\vec P + \vec k}); \label{29}
	\\
	|N_{\vec P}|^2 &= \phi_{\vec P}^2.\label{30}
\end{align}

From here we can immediately conclude that the quantity $G$ yields only a correction of the order of $g^4$ and can be neglected. Let us expand the Fourier transform of the function $\phi(\vec r)$ into Taylor series over $\vec P$ up to second-order
\begin{align}
	\phi^2_{\vec P} &=\phi_0^2 e^{-\frac{P^2}{\lambda^2}} \approx \phi_0^2 \left(1-\frac{P^2}{\lambda^2}\right), \label{31}
	\\
	\phi_{\vec P - \vec k}^2 &= \phi_0^2 e^{-\frac{(\vec P - \vec k)^2}{\lambda^2}} = \phi_0^2\left(e^{-\frac{k^2}{\lambda^2}}+e^{-\frac{k^2}{\lambda^2}}\frac{2\thp{P}{k}}{\lambda^2}+e^{-\frac{k^2}{\lambda^2}}\frac{2(\thp{P}{k})^2}{\lambda^4}-e^{-\frac{k^2}{\lambda^2}} \frac{P^2}{\lambda^2}\right),\label{32}
	\\
	\phi_{\vec P - \vec k} &= \phi_0 e^{-\frac{(\vec P - \vec k)^2}{2\lambda^2}} = \phi_0\left(e^{-\frac{k^2}{2\lambda^2}}+e^{-\frac{k^2}{2\lambda^2}}\frac{\thp{P}{k}}{\lambda^2}+e^{-\frac{k^2}{2\lambda^2}}\frac{(\thp{P}{k})^2}{2\lambda^4}-e^{-\frac{k^2}{2\lambda^2}} \frac{P^2}{2\lambda^2}\right).\label{33}
\end{align}

By plugging equations (\ref{31}-\ref{33}) into equations (\ref{26}-\ref{30}) and taking into account only the terms of the order of $P^2$, we find for the vector $\vec Q$
\begin{align}
	\vec Q &= \frac{1}{1-P^2/\lambda^2}\frac{2}{\lambda^2}\sum_{\vec k}\vec k |u_{\vec k}|^2 e^{-\frac{k^2}{\lambda^2}}(\thp{P}{k}) = \frac{1}{1-P^2/\lambda^2}\frac{g^2}{\lambda^2}\frac{\vec P}{4\pi^2}\int_0^\infty dk k e^{-\frac{3}{2}\frac{k^2}{\lambda^2}}\int_{-1}^1 t^2 dt \nonumber
	\\
	&= \frac{g^2}{9\pi^2}\frac{\vec P}{2} \label{34}
\end{align}
and for the field energy
\begin{align}
	E_{\text{f}} = \frac{1}{1-P^2/\lambda^2} \sum_{\vec k}\left(k + \frac{k^2}{2}\right)|u_{\vec k}|^2 \left(e^{-\frac{k^2}{\lambda^2}}+e^{-\frac{k^2}{\lambda^2}}\frac{2\thp{P}{k}}{\lambda^2}+e^{-\frac{k^2}{\lambda^2}}\frac{2(\thp{P}{k})^2}{\lambda^4}-e^{-\frac{k^2}{\lambda^2}} \frac{P^2}{\lambda^2}\right). \label{35}
\end{align}
In this expression, in round brackets the first and the last terms cancel each other after decomposition of the normalization constant in the Taylor series in $\vec P$. The result reads as
\begin{align}
	E_{\text{f}} &= E_{\text{f}}(0)+\frac{2}{\lambda^2}\sum_{\vec k}\left(k + \frac{k^2}{2}\right)|u_{\vec k}|^2e^{-\frac{k^2}{\lambda^2}}(\thp{P}{k})^2 \nonumber
	\\
	&=E_{\text{f}}(0)+\frac{g^2}{\lambda^4}\frac{P^2}{4\pi^2}\int_0^\infty k \left(k + \frac{k^2}{2}\right)e^{-\frac{3}{2}\frac{k^2}{\lambda^2}}dk\int_{-1}^1 t^2 dt \nonumber
	\\
&=E_{\text{f}}(0) + \frac{P^2}{2}\frac{g^2}{9\pi^2}\frac{1}{6\lambda}(\sqrt{6\pi}+2\lambda). \label{36}
\end{align}

The remaining energy is calculated in exactly the same way
\begin{align}
	E_{\text{int}} = \frac{g}{\sqrt{2 \Omega}}\frac{1}{1-P^2/(2\lambda^2)}\sum_{\vec k}\frac{u_{\vec k}}{\sqrt{k}}\Bigg[e^{-\frac{k^2}{2\lambda^2}}+e^{-\frac{k^2}{2\lambda^2}}\frac{\thp{P}{k}}{\lambda^2}+e^{-\frac{k^2}{2\lambda^2}}\frac{(\thp{P}{k})^2}{2\lambda^4}-e^{-\frac{k^2}{2\lambda^2}} \frac{P^2}{2\lambda^2} \nonumber
	\\
	+e^{-\frac{k^2}{2\lambda^2}}-e^{-\frac{k^2}{2\lambda^2}}\frac{\thp{P}{k}}{\lambda^2}+e^{-\frac{k^2}{2\lambda^2}}\frac{(\thp{P}{k})^2}{2\lambda^4}-e^{-\frac{k^2}{2\lambda^2}} \frac{P^2}{2\lambda^2}\Bigg]. \label{37}
\end{align}
In a full analogy to the field energy $E_{\text{f}}$, the first and the last terms are cancelled. The remaining terms are
\begin{align}
	E_{\text{int}} &= E_{\text{int}}(0) + \frac{g}{\sqrt{2 \Omega}}\frac{1}{\lambda^4} \sum_{\vec k}\frac{u_{\vec k}}{\sqrt{k}}e^{-\frac{k^2}{2\lambda^2}}(\thp{P}{k})^2 \nonumber
	\\
	&=E_{\text{int}}(0) -\frac{g^2}{\lambda^4}\frac{P^2}{8\pi^2}\int_0^\infty k^2 e^{-\frac{3}{4}\frac{k^2}{\lambda^2}}dk\int_{-1}^1 t^2 dt \nonumber
	\\
	&=E_{\text{int}}(0) - \frac{g^2}{9\pi^2}\frac{1}{\lambda}\sqrt{\frac{\pi}{3}}\frac{P^2}{2}. \label{38}
\end{align}

By combining all results together, we find the equation for the total energy of the system  with a renormalized mass, in the zeroth-order approximation:
\begin{align}
	E^{(0)}_L(\vec P, g) &= E^{(0)}_L(0, g) + \frac{P^2}{2}\left[1-\frac{g^2}{9\pi^2}\left(1+\frac{1}{\lambda}\sqrt{\frac{\pi}{3}} - \frac{(\sqrt{6\pi}+2\lambda)}{6\lambda}\right)\right] \nonumber
	\\
	&=E^{(0)}_L(0, g) + \frac{P^2}{2}\left[1-\frac{g^2}{9\pi^2}\left(\frac{2}{3}+\frac{1}{\lambda}\frac{\sqrt{6\pi}(\sqrt{2}-1)}{6}\right)\right], \label{39}
\end{align}
or by plugging in for $\lambda$ according to equation (\ref{23}) we finally obtain
\begin{align}
	E^{(0)}_L(\vec P, g) = E^{(0)}_L(0, g) + \frac{P^2}{2}\left[1-\frac{g^2}{9\pi^2}\frac{17-\sqrt{2}}{21}\right]. \label{40}
\end{align}

Concluding, the renormalized mass is equal to
\begin{align}
	m^{*(0)} = 1+\frac{g^2}{9\pi^2}\frac{17-\sqrt{2}}{21}. \label{41}
\end{align}
\section*{Appendix E: Calculation of matrix elements in the second iteration for the energy of the system} 
\label{sec:calculation_of_matrix_elements_in_the_second_iteration_for_the_energy_of_the_system}
The calculation of the second iteration of the energy of the system requires the evaluation of the transition matrix elements $\la\Psi^{(L)}_{\vec P}| \opa H |\Psi_{\vec P_1, \{ n_{\vec k}\}}\ra$, $\la\Psi^{(L)}_{\vec P}| \Psi_{\vec P_1,   \{ n_{\vec k}\}}\ra$ and $\la \Psi_{\vec P_1,   \{ n_{\vec k}\}}| \opa H |\Psi_{\vec P_1,   \{ n_{\vec k}\}}\ra$ from the full Hamiltonian of the system, equation (\ref{7}), with the function
\begin{align}
	|\Psi^{(0)}_{\vec P_1,    n_{\vec k}}\ra &=  \frac{1}{N_{\vec P_1,1_{\vec k}}\sqrt{\Omega}}\int d \vec R \phi_{\vec P_1}(\vec r- \vec R) \exp\left\{\ri (\vec P_1 - \vec k n_{\vec k})\cdot \vec R\right\}\nonumber
	\\
	&\times\exp\left[\sum_{\vec k}(u_{\vec k} e^{-\ri\thp{k}{R}}\opa a_{\vec k}^\dag - u_{\vec k}^* e^{\ri\thp{k}{R}}\opa a_{\vec k})\right]|n_{\vec k}\ra \nonumber
	\\
	&=\frac{1}{N_{\vec P_1,1_{\vec k}}\sqrt{\Omega}}\int d \vec R \phi_{\vec P_1}(\vec r- \vec R) \exp\left\{\ri (\vec P_1 - \vec k n_{\vec k})\cdot \vec R - \frac{1}{2}\sum_{{\vec k}}|u_{\vec k}|^2 + \sum_{\vec k} u_{\vec k} e^{-\ri\thp{k}{R}}\opa a_{\vec k}^\dag \right\}\nonumber
	\\
	&\times\frac{(\opa a_{\vec k}^\dag - u_{\vec k}^* e^{\ri\thp{k}{R}})^{n_{\vec k}}}{\sqrt{n_{\vec k}!}}|0\ra. \label{42}
\end{align}

The normalization constant in equation (\ref{42}) is calculated with the help of equation (\ref{4}) and has the form
\begin{align}
	|N_{\vec P_1,1_{\vec k}}|^2 &= \int d \vec R_1 d \vec \rho \phi_{\vec P_1}^*(\vec \rho)\phi_{\vec P_1}(\vec \rho - \vec R_1) e^{\ri(\vec P_1 - \vec k)\cdot \vec R_1 + \sum_{\vec k} |u_{\vec k}|^2(e^{-\ri\thp{k}{R_1}}-1)}\nonumber
	\\
	&\times\left(2|u_{\vec k}|^2(\cos\thp{k}{R_1}-1)+1\right).
\end{align}

The calculation of the transition matrix element is performed with the help of equation (\ref{4}):
\begin{align}
	\la \Psi_{\vec P_1,n_{\vec k}}&|\opa H|\Psi_{\vec P}^{(L)}\ra = \frac{(2\pi)^3 \delta(\vec P - \vec P_1)}{N_{\vec P_1,1_{\vec k}}^* N_{\vec P} \Omega} \int d \vec R_1 d \vec \rho \phi_{\vec P_1}^*(\vec \rho)\phi_{\vec P}(\vec \rho - \vec R_1) \nonumber
	\\
	&\times e^{\ri\thp{P}{R_1} + \sum_{\vec k} |u_{\vec k}|^2(e^{-\ri\thp{k}{R_1}}-1)}\frac{u_{\vec k}^{n_{\vec k}} (e^{-\ri\thp{k}{R_1}}-1)^{n_{\vec k}}}{\sqrt{n_{\vec k}!}} \nonumber
	\\
	&\times\Bigg[\frac{P_1^2}{2} + \left(\frac{1}{2}k^2+k- \thp{P_1}{k}\right) n_{\vec k} e^{-\ri\thp{k}{R_1}}(e^{-\ri\thp{k}{R_1}}-1)^{-1} \nonumber
	\\
	&\mspace{35mu}+\frac{g}{\sqrt{2 \Omega}}\frac{e^{-\ri\thp{k}{\rho}} }{\sqrt{\omega_k}}n_{\vec k} u_{\vec k}^{-1}(e^{-\ri\thp{k}{R_1}}-1)^{-1} \nonumber
	\\
	&\mspace{35mu}+\sum_{\vec m} \vec m |u_{\vec m}|^2 e^{-\ri\thp{m}{R_1}}\cdot \left(-\vec P_1+\vec k n_{\vec k} e^{-\ri\thp{k}{R_1}}(e^{-\ri\thp{k}{R_1}}-1)^{-1}\right) \nonumber
	\\
	&\mspace{35mu}+ \sum_{\vec m} \left(\frac{1}{2}m^2+m\right) |u_{\vec m}|^2 e^{-\ri\thp{m}{R_1}}+ \frac{1}{2}k^2 n_{\vec k}(n_{\vec k}-1)e^{-2\ri\thp{k}{R_1}}(e^{-\ri\thp{k}{R_1}}-1)^{-2} \nonumber
	\\
	&\mspace{35mu}+\frac{1}{2}\left(\sum_{\vec m} \vec m |u_{\vec m}|^2 e^{-\ri\thp{m}{R_1}}\right)^2+\frac{g}{\sqrt{2 \Omega}}\sum_{\vec m} \frac{u_{\vec m} e^{\ri\thp{m}{\rho}}}{\sqrt{\omega_m}}(e^{-\ri\thp{m}{R_1}}+1)\Bigg]. \label{44}
\end{align}
We further obtain the cover integral
\begin{align}
	\la \Psi_{\vec P_1,n_{\vec k}}|\Psi_{\vec P}^{(L)}\ra &= \frac{(2\pi)^3 \delta(\vec P - \vec P_1)}{N_{\vec P_1,1_{\vec k}}^* N_{\vec P} \Omega} \int d \vec R_1 d \vec \rho \phi_{\vec P_1}^*(\vec \rho)\phi_{\vec P}(\vec \rho - \vec R_1) e^{\ri\thp{P}{R_1} + \sum_{\vec k} |u_{\vec k}|^2(e^{-\ri\thp{k}{R_1}}-1)}\nonumber
	\\
	&\times\frac{u_{\vec k}^{n_{\vec k}} (e^{-\ri\thp{k}{R_1}}-1)^{n_{\vec k}}}{\sqrt{n_{\vec k}!}} \label{45}
\end{align}
and the expectation value of the Hamiltonian
\begin{align}
	\la \Psi_{\vec P_1,1_{\vec k}}|\opa H|\Psi_{\vec P_1,1_{\vec k}}\ra &= \frac{P_1^2}{2} + \frac{1}{|N_{\vec P_1,1_{\vec k}}|^2 } \int d \vec R_1 d \vec \rho \phi_{\vec P_1}^*(\vec \rho)\phi_{\vec P_1}(\vec \rho - \vec R_1) e^{\ri(\vec P_1 - \vec k)\cdot \vec R_1 + \sum_{\vec k} |u_{\vec k}|^2(e^{-\ri\thp{k}{R_1}}-1)} \nonumber
	\\
	&\mspace{-105mu}\times\Bigg\{k^2 |u_{\vec k}|^2e^{-\ri\thp{k}{R_1}}\nonumber
	\\
	&\mspace{-75mu}+\left(2|u_{\vec k}|^2(\cos\thp{k}{R_1}-1)+1\right)\Bigg[ - \sum_{\vec m} \thp{P_1}{m} |u_{\vec m}|^2 e^{-\ri\thp{m}{R_1}} + \sum_{\vec m} \left(\frac{m^2}{2}+m\right)|u_{\vec m}|^2 e^{-\ri\thp{m}{R_1}}\nonumber
	\\
	&\mspace{90mu}+\frac{1}{2}\left(\sum_{\vec m} \vec m |u_{\vec m}|^2 e^{-\ri\thp{m}{R_1}}\right)^2 + \frac{g}{\sqrt{2 \Omega}}\sum_{\vec m} \frac{u_{\vec m} e^{\ri\thp{m}{\rho}}}{\sqrt{\omega_m}}(e^{-\ri\thp{m}{R_1}}+1)\Bigg]\nonumber
	\\
	&\mspace{-75mu}+\left(2|u_{\vec k}|^2(e^{-\ri\thp{k}{R_1}}-1)+1\right)\Bigg[-\thp{k}{P_1} + \sum_{\vec m} \thp{k}{m}|u_{\vec m}|^2 e^{-\ri\thp{m}{R_1}}+\frac{k^2}{2}+k\Bigg]\nonumber
	\\
	&\mspace{-75mu}+\frac{g}{\sqrt{2 \Omega}}\Bigg[\frac{u_{\vec k} e^{\ri\thp{k}{\rho}}}{\sqrt{\omega_{\vec k}}}(e^{-\ri\thp{k}{R_1}}-1) + \frac{u_{\vec k}^* e^{-\ri\thp{k}{\rho}}}{\sqrt{\omega_{\vec k}}}(1-e^{\ri\thp{k}{R_1}})\Bigg]\Bigg\}. \label{46}
\end{align}

Equations (\ref{44}-\ref{46}) are valid for arbitrary coupling constants. However, in the weak coupling limit they are significantly simplified as they can be expressed via Fourier transforms. Another significant simplification comes from the fact that the action of one field mode on the system is inversely proportional to the square root of the normalization volume $\Omega$, that is $u_{\vec k}\sim 1/\sqrt{\Omega}$. Consequently, such terms can be kept only within the sum. Within this approximation, equations (\ref{44}-\ref{46}) can be rewritten as
\begin{align}
	\la \Psi_{\vec P_1,n_k}|\Psi_{\vec P}^{(L)}\ra = \frac{(2\pi)^3 \delta(\vec P - \vec P_1)}{\Omega}\frac{u_{\vec k}(\phi_{\vec P - \vec k}^2- \phi_{\vec P}^2)}{\phi_{\vec P} \phi_{\vec P - \vec k}},\label{47}
\end{align}
and
\begin{align}
		\la \Psi_{\vec P_1,n_k}|\opa H|\Psi_{\vec P}^{(L)}\ra &= \frac{(2\pi)^3 \delta(\vec P - \vec P_1)}{\Omega}\frac{1}{\phi_{\vec P}\phi_{\vec P - \vec k}} \nonumber
		\\
		&\times \Bigg[\frac{P_1^2}{2} u_{\vec k}(\phi_{\vec P - \vec k}^2- \phi_{\vec P}^2) + \left(\frac{k^2}{2}+k- \thp{P_1}{k}\right)u_{\vec k} \phi_{\vec P - \vec k}^2+\frac{g}{\sqrt{2 \Omega}}\frac{\phi_{\vec P - \vec k} \phi_{\vec P}}{\sqrt{k}}\nonumber
		\\
		&-u_{\vec k}(\vec P_1 - \vec k)\sum_{\vec m}\vec m |u_{\vec m}|^2 \phi_{\vec P - \vec m - \vec k}^2  \nonumber
		\\
		&+ u_{\vec k}\sum_{\vec m}\left(\frac{m^2}{2}+m\right)|u_{\vec m}|^2 \phi_{\vec P - \vec m - \vec k}^2 +\frac{u_{\vec k}}{2}\sum_{\vec l,\vec m}\thp{l}{m}|u_{\vec l}|^2|u_{\vec m}|^2 \phi_{\vec P - \vec l - \vec m - \vec k}^2 \nonumber
		\\
		&+\frac{g}{\sqrt{2 \Omega}}u_{\vec k}\sum_{\vec m}\frac{u_{\vec m}}{\sqrt{m}}\phi_{\vec P - \vec k}(\phi_{\vec P - \vec k - \vec m}+\phi_{\vec P - \vec k + \vec m}) - u_{\vec k}\phi_{\vec P}^2 \tilde E_L^{(0)}\Bigg], \label{48}
\end{align}
as well as
\begin{align}
	\la \Psi_{\vec P_1,n_k}|\opa H|\Psi_{\vec P_1,n_k}\ra &= \frac{P_1^2}{2}+\frac{1}{\phi_{\vec P - \vec k}^2}\Bigg[\phi_{\vec P -\vec k}^2\left(-\thp{k}{P_1}+\frac{k^2}{2}+k\right) - (\vec P_1-\vec k)\cdot\sum_{\vec m}\vec m |u_{\vec m}|^2\phi_{\vec P - \vec m - \vec k}^2\nonumber
	\\
	&+\sum_{\vec m}\left(\frac{m^2}{2}+m\right)|u_{\vec m}|^2\phi_{\vec P - \vec m - \vec k}^2 +\frac{g}{\sqrt{2 \Omega}}\sum_{\vec m}\frac{u_{\vec m}}{\sqrt{m}}\phi_{\vec P - \vec k}\left(\phi_{\vec P - \vec m-\vec k}+\phi_{\vec P + \vec m-\vec k}\right)\nonumber
	\\
	&+\frac{1}{2}\sum_{\vec l,\vec m}\thp{l}{m}|u_{\vec l}|^2|u_{\vec m}|^2 \phi_{\vec P - \vec m  - \vec l - \vec k}^2 \Bigg]. \label{49}
\end{align}

In equations (\ref{47}-\ref{49}) we have used expressions for the normalization constants
\begin{align}
	N_{\vec P} = \phi_{\vec P}, \quad N_{\vec P_1,1_{\vec k}} = \phi_{\vec P_1 - \vec k} \label{50}
\end{align}
and sorted out the energy of the zeroth-order approximation
\begin{align}
	\tilde E_L^{(0)} &= - \vec P\sum_{\vec m}\vec m |u_{\vec m}|^2 \frac{\phi_{\vec P - \vec m}^2}{\phi_{\vec P}^2} + \frac{1}{2}\sum_{\vec l,\vec m}\thp{m}{l}|u_{\vec m}|^2 |u_{\vec l}|^2 \frac{\phi_{\vec P - \vec m - \vec l}^2}{\phi_{\vec P}^2} \nonumber
	\\
	&+\sum_{\vec m}\left(\frac{m^2}{2} + m\right)|u_{\vec m}|^2 \frac{\phi_{\vec P - \vec m}^2}{\phi_{\vec P}^2} + \frac{g}{\sqrt{2 \Omega}}\sum_{\vec m}\frac{u_{\vec m}}{\sqrt{m}}\phi_{\vec P}\frac{(\phi_{\vec P - \vec m}+\phi_{\vec P + \vec m})}{\phi_{\vec P}}. \label{51}
\end{align}
\section*{Appendix F: Second iteration for the energy of the system with particle at rest} 
\label{sec:second_iteration_for_the_energy_of_the_system_with_particle_at_rest}
For the evaluation of the second iteration for the particle energy the knowledge of the behavior of different terms in expressions (\ref{47}-\ref{49}) is required. In order to determine those, we will carry out the summations over $\vec m$. For the first sum we can write
\begin{align}
	\vec I^{(1)}_{\vec k} &= \frac{1}{g^2\phi_{\vec k}^2}\sum_{\vec m}\vec m |u_{\vec m}|^2 \phi_{\vec m+\vec k}^2 =\frac{\phi_0^2}{g^2\phi_{\vec k}^2} \frac{g^2}{2(2\pi)^3}\int d\vec m \left(\begin{aligned}
		&m\sin \theta \cos \phi
		\\
		&m\sin \theta \sin \phi
		\\
		&m \cos \theta
	\end{aligned}\right)\frac{e^{-\frac{m^2}{2 \lambda^2}}}{m^3}e^{-\frac{m^2 + k^2 + 2\thp{m}{k}}{\lambda^2}} \nonumber
	\\
	&=\frac{1}{8\pi^2}\frac{\vec k}{k}\int dm e^{-\frac{3}{2}\frac{m^2}{\lambda^2}}\frac{-2km \lambda^2 \cosh \frac{2km}{\lambda^2} + \lambda^4 \sinh \frac{2km}{\lambda^2}}{2k^2 m^2}\nonumber
	\\
	&=\frac{\lambda^2}{32\pi^2}\frac{\vec k}{k}\frac{4k-e^{\frac{2}{3}\frac{k^2}{\lambda^2}}\sqrt{6\pi}\lambda \text{Erf}\frac{\sqrt{\frac{2}{3}}k}{\lambda}}{k^2} \underset{k\rightarrow\infty}{\longrightarrow}-\frac{\lambda^2}{32\pi^2}\frac{\vec k}{k^3}\left(-4k+e^{\frac{2}{3}\frac{k^2}{\lambda^2}}\sqrt{6\pi}\lambda\right) \nonumber
	\\
	&\sim - \frac{\phi_0^2}{\phi_{\vec k}^2} \frac{\lambda^3 \sqrt{6\pi}}{32\pi^2}\frac{e^{-\frac{1}{3}\frac{k^2}{\lambda^2}}}{k^3}\vec k \label{52}
\end{align}
as for the second
\begin{align}
	I^{(2)}_{\vec k} &= \frac{1}{g^2 \phi_{\vec k}^2}\sum_{\vec m}\left(\frac{m^2}{2}+m\right)|u_{\vec m}|^2 \phi_{\vec m+\vec k}^2 = \frac{1}{2(2\pi)^3}\int \frac{d \vec m}{m^2}\left(1+\frac{m}{2}\right)e^{-\frac{3}{2}\frac{m^2}{\lambda^2}-\frac{2\thp{m}{k}}{\lambda^2}} \nonumber
	\\
	&=\frac{\lambda^2}{8\pi^2}\int dm\left(1+\frac{m}{2}\right)e^{-\frac{3}{2}\frac{m^2}{\lambda^2}} \frac{\sinh \frac{2km}{\lambda^2}}{km}\nonumber
	\\
	&=\frac{\lambda^2}{96k}\left(\frac{\sqrt{6\pi}\lambda e^{\frac{2}{3}\frac{k^2}{\lambda^2}}\text{Erf}\frac{\sqrt{\frac{2}{3}}k}{\lambda}}{\pi^2}+\frac{6\text{Erfi}\frac{\sqrt{\frac{2}{3}}k}{\lambda}}{\pi}\right) \nonumber
	\\
	&\underset{k\rightarrow\infty}{\longrightarrow} \frac{\lambda^3}{96k^2 \pi^{\frac{3}{2}}}\left(\sqrt{6}e^{\frac{2}{3}\frac{k^2}{\lambda^2}}(3+k)-\frac{6\ri k\sqrt{\pi}}{\lambda}\right) \sim \frac{\phi_0^2}{\phi_{\vec k}^2} \frac{\lambda^3\sqrt{6\pi}}{96\pi^2}\frac{e^{-\frac{1}{3}\frac{k^2}{\lambda^2}}}{k} \label{53}
\end{align}
and for the third
\begin{align}
	I^{(3)}_{\vec k} &= \frac{1}{g^2 \phi_{\vec k}^2}\phi_{\vec k}\frac{g}{\sqrt{2 \Omega}}\sum_{\vec m}\frac{u_{\vec m}}{\sqrt{m}}(\phi_{\vec m+\vec k}+\phi_{\vec m - \vec k}) = - \frac{1}{2(2\pi)^3}\int \frac{d\vec m}{m^2}e^{-\frac{3}{4}\frac{m^2}{\lambda^2}}\left(e^{-\frac{\thp{m}{k}}{\lambda^2}}+e^{\frac{\thp{m}{k}}{\lambda^2}}\right) \nonumber
	\\
	&= -\frac{\lambda^2}{2\pi^2}\int dm e^{-\frac{3}{4}\frac{m^2}{\lambda^2}}\frac{\sinh\frac{km}{\lambda^2}}{km} = -\frac{\lambda^2}{4\pi}\frac{\text{Erfi}\frac{k}{\sqrt{3}\lambda}}{k} \nonumber
	\\
	&\underset{k\rightarrow\infty}{\longrightarrow} -\frac{\lambda^2}{4k^2 \pi^{\frac{3}{2}}}(-\ri k \sqrt{\pi}+\sqrt{3}\lambda e^{\frac{k^2}{3 \lambda^2}}) \sim -\frac{\phi_0^2}{\phi_{\vec k}^2}\frac{\lambda^3 \sqrt{3\pi}}{4\pi^2}\frac{e^{-\frac{2}{3}\frac{k^2}{\lambda^2}}}{k^2}. \label{54}
\end{align}

In the above expressions $\text{Erf}(x) = 2/\sqrt{\pi}\int_0^x e^{-z^2}dz$ and $\text{Erfi}(x) = -\ri\text{Erf}(\ri x)$ are the error function and the imaginary error functions, respectively.

Consequently, we can introduce the abbreviations, which where used in equations (\ref{eq:second_order_iteration_for_the_energy_convergence_4}-\ref{eq:second_order_iteration_for_the_energy_convergence_8}) of the manuscript, namely
\begin{align}
	I_{\vec k} &= \vec k \cdot \vec I_{\vec k}^{(1)}+I_{\vec k}^{(2)}+I_{\vec k}^{(3)}\underset{k\rightarrow\infty}\sim-\frac{\lambda^3\sqrt{6\pi}}{48\pi^2}\frac{e^{\frac{2k^2}{3 \lambda^2}}}{k}. \label{55}
\end{align}

After the determination of the asymptotic behavior of the different terms, we can find the second iteration for the energy of the system
\begin{align}
	E^{(2)} = \frac{A}{B}. \label{56}
\end{align}
For the numerator we have
\begin{align}
	A = E_L^{(0)} &+ \sum_{\vec P_1, \{n_k\}}C^{(1)}_{\vec P_1, \{n_{\vec k}\}}\la\Psi^{(L)}_{\vec P}| \opa H |\Psi_{\vec P_1,   \{ n_{\vec k}\}}\ra \nonumber
	\\
	&=E_L^{(0)} + \sum_{\vec k<\vec k_0}\frac{-\left(u_{\vec k} \frac{\phi_{\vec k}}{\phi_0}\left(\frac{k^2}{2}+k\right)+\frac{g}{\sqrt{2 \Omega}}\frac{1}{\sqrt{k}}\right)^2}{(\frac{k^2}{2}+k)}+\sum_{\vec k>\vec k_0}\frac{(-\frac{g}{\sqrt{2 \Omega}}\frac{\phi_{\vec k} \phi_0}{\sqrt{k}})(-E_0 u_{\vec k})}{\phi_{\vec k}^2 I_{\vec k}g^2} \label{57}
\end{align}
and for the denominator we obtain
\begin{align}
	B = 1 &+ \sum_{\vec P_1, \{n_k\}}C^{(1)}_{\vec P_1, \{n_{\vec k}\}}\la\Psi^{(L)}_{\vec P}| \Psi_{\vec P_1,   \{ n_{\vec k}\}}\ra \nonumber
	\\
	&=1+ \sum_{\vec k<\vec k_0}\frac{-\left(u_{\vec k} \frac{\phi_{\vec k}^2}{\phi_0^2}\left(\frac{k^2}{2}+k\right)+\frac{g}{\sqrt{2 \Omega}}\frac{\phi_{\vec k}}{\phi_0\sqrt{k}}\right)u_{\vec k}(1-\frac{\phi_0^2}{\phi_{\vec k}^2})}{(\frac{k^2}{2}+k)}+ \sum_{\vec k>\vec k_0}\frac{(-\frac{g}{\sqrt{2 \Omega}}\frac{\phi_{\vec k} \phi_0}{\sqrt{k}})u_{\vec k}(\frac{\phi_{\vec k}^2}{\phi_0^2}-1)}{\phi_{\vec k}^2 I_{\vec k}g^2 }. \label{58}
\end{align}

Further explicit calculations yield
\begin{align}
	A = E_L^{(0)} &+ \sum_{\vec k<\vec k_0}\frac{-\left(u_{\vec k} \frac{\phi_{\vec k}}{\phi_0}\left(\frac{k^2}{2}+k\right)+\frac{g}{\sqrt{2 \Omega}}\frac{1}{\sqrt{k}}\right)^2}{(\frac{k^2}{2}+k)} +\sum_{\vec k>\vec k_0}\frac{(-\frac{g}{\sqrt{2 \Omega}}\frac{\phi_{\vec k} \phi_0}{\sqrt{k}})(-E_0 u_{\vec k})}{\phi_{\vec k}^2 I_{\vec k}g^2 } \nonumber
	\\
	&=E_L^{(0)}-\sum_{\vec k<\vec k_0}\left(u_{\vec k}^2\left(\frac{\phi_{\vec k}}{\phi_0}\right)^2\left(\frac{k^2}{2}+k\right) +\frac{2g}{\sqrt{2 \Omega}}\frac{u_{\vec k}}{\sqrt{k}}\frac{\phi_{\vec k}}{\phi_0}+\frac{g^2}{2 \Omega}\frac{1}{k^2(k/2+1)}\right)\nonumber
	\\
	&+\frac{g}{\sqrt{2 \Omega}}\frac{E_L^{(0)}}{g^2} \sum_{\vec k>\vec k_0}\frac{\phi_{\vec k}\phi_0 u_{\vec k}}{\sqrt{k}\phi_{\vec k}^2 I_{\vec k}} \nonumber
	\\
	&=E_L^{(0)}-\sum_{\vec k<\vec k_0}\left(u_{\vec k}^2\left(\frac{\phi_{\vec k}}{\phi_0}\right)^2\left(\frac{k^2}{2}+k\right) +\frac{2g}{\sqrt{2 \Omega}}\frac{u_{\vec k}}{\sqrt{k}}\frac{\phi_{\vec k}}{\phi_0}\right)-\frac{g^2}{2\pi^2}\ln\left(\frac{k_0}{2}+1\right)\nonumber
	\\
	&+\frac{E_L^{(0)}}{2(2\pi)^3} \int_{\vec k_0}^\infty d\vec k \frac{e^{-\frac{k^2}{4 \lambda^2}}}{k^2}\frac{e^{-\frac{k^2}{2 \lambda^2}}}{\frac{\lambda^3 \sqrt{6\pi}}{48\pi^2}\frac{e^{-\frac{1}{3}\frac{k^2}{\lambda^2}}}{k}} \nonumber
	\\
	&=E_L^{(0)}-\sum_{\vec k<\vec k_0}\left(u_{\vec k}^2\left(\frac{\phi_{\vec k}}{\phi_0}\right)^2\left(\frac{k^2}{2}+k\right) +\frac{2g}{\sqrt{2 \Omega}}\frac{u_{\vec k}}{\sqrt{k}}\frac{\phi_{\vec k}}{\phi_0}\right)-\frac{g^2}{2\pi^2}\ln\left(\frac{k_0}{2}+1\right)\nonumber
	\\
	&+E_{L}^{(0)}\frac{12}{\lambda^3 \sqrt{6\pi}} \int_{k_0}^\infty dk k e^{-\frac{5}{12}\frac{k^2}{\lambda^2}} \nonumber
	\\
	&=E_L^{(0)}-\left[\frac{g^2\lambda}{24 \pi ^2}\left(\sqrt{6 \pi } \text{Erf}\left(\frac{\sqrt{\frac{3}{2}} k_0}{\lambda }\right)+\lambda -\lambda e^{-\frac{3k_0^2}{2 \lambda ^2}}\right)-\frac{g^2\lambda}{2 \sqrt{3}\pi^{3/2}}\text{Erf}\left(\frac{\sqrt{3}k_0}{2\lambda }\right)\right]\nonumber
	\\
	&-\frac{g^2}{2\pi^2}\ln\left(\frac{k_0}{2}+1\right)+E_L^{(0)}\frac{12\sqrt{6\pi}}{5 \lambda \pi}e^{-\frac{5k_0^2}{12 \lambda^2}} \label{59}
\end{align}
and
\begin{align}
	B &= 1+ \sum_{\vec k<\vec k_0}\frac{-\left(u_{\vec k} \frac{\phi_{\vec k}^2}{\phi_0^2}\left(\frac{k^2}{2}+k\right)+\frac{g}{\sqrt{2 \Omega}}\frac{\phi_{\vec k}}{\phi_0\sqrt{k}}\right)u_{\vec k}(1-\frac{\phi_0^2}{\phi_{\vec k}^2})}{(\frac{k^2}{2}+k)} + \sum_{\vec k>\vec k_0}\frac{(-\frac{g}{\sqrt{2 \Omega}}\frac{\phi_{\vec k} \phi_0}{\sqrt{k}})u_{\vec k}(\frac{\phi_{\vec k}^2}{\phi_0^2}-1)}{\phi_{\vec k}^2 I_{\vec k}g^2}\nonumber
	\\
	&=1+\frac{g^2}{2(2\pi)^3}\int_0^{\vec k_0} d\vec k \frac{k^2}{\lambda^2}\left[\frac{e^{-\frac{3k^2}{2 \lambda^2}}}{k^3}-\frac{e^{-\frac{3k^2}{4 \lambda^2}}}{k^2(k^2/2+k)}\right]-\frac{12}{\lambda^5 \sqrt{6\pi}}\int_{k_0}^\infty dk k^3 e^{-\frac{5}{12}\frac{k^2}{\lambda^2}} \nonumber
	\\
	&=1+\frac{g^2}{12\pi^2}(1-e^{-\frac{3}{2}\frac{k_0^2}{\lambda^2}})-g^2f\left(\frac{k_0}{\lambda}\right)- \frac{144\sqrt{6\pi}}{25 \lambda \pi}\left(1+\frac{5}{12}\frac{k_0^2}{\lambda^2}\right)e^{-\frac{5k_0^2}{12 \lambda^2}} \label{60}
\end{align}
with
\begin{align*}
	f(x) = \frac{1}{4\pi^2}\int_0^x\frac{tdt}{1+t/2}e^{-\frac{3}{4}t^2}.
\end{align*}

\bibliography{quantum-2}

\end{document}